\documentclass[aps,prl,reprint,showpacs,noshowkeys,floatfix]{revtex4-1}

\usepackage{graphicx}

\usepackage{amsmath}
\usepackage{mathtools}
\usepackage{amssymb}
\usepackage{amsfonts}
\usepackage{mathrsfs}

\usepackage[backref=none]{hyperref}
\usepackage[usenames,dvipsnames]{color}
\definecolor{MyDarkBlue}{rgb}{0,0.1,0.7}
\hypersetup{pdfborder={0 0 0},colorlinks,breaklinks=true,
  urlcolor={MyDarkBlue},citecolor={MyDarkBlue},linkcolor={MyDarkBlue}
}

\newcommand{\rmi}{\mathrm{i}}
\newcommand{\rme}{\mathrm{e}}
\newcommand{\rmd}{\mathrm{d}}
\newcommand{\rmf}{\mathrm{f}}

\newcommand{\bq}{{\bf q}}
\newcommand{\bx}{{\bf x}}
\newcommand{\nb}{\bar{\nu}}
\newcommand{\bqi}{{\bf q}^{\rm i}}
\newcommand{\bqf}{{\bf q}^{\rm f}}
\newcommand{\bxf}{{\bf x}^{\rm f}}
\newcommand{\bxi}{{\bf x}^{\rm i}}
\newcommand{\blam}{\boldsymbol{\lambda}}
\newcommand{\lamT}{\lambda_{\rm T}}

\DeclareMathOperator*{\Linv}{\mathscr{L}^{-1}_\beta\!}
\DeclareMathOperator*{\Linvs}{\mathscr{L}^{-1}_s\!}
\DeclareMathOperator*{\atan}{{\rm tan}^{-1}}
\DeclareMathOperator*{\erfc}{{\rm erfc}}

\newcommand{\eg}{\textit{e.g.}\ }

\newcommand{\wrt}{\textit{w.r.t.}\ }

\newcommand{\eref}[1]{(\ref{#1})}
\newcommand{\Leref}[1]{Eq.~(\ref*{#1}) of the main text}

\newcommand{\fref}[1]{Fig.~\ref{#1}}

\newcommand{\Q}[1]{}

\begin{document}

\title{Canonical description of 1D few-body systems with short range interaction}
\date{\today}

\author{Quirin Hummel}
\email{quirin.hummel@ur.de}
\author{Juan Diego Urbina}
\author{Klaus Richter}
\affiliation{Institut f\"ur Theoretische Physik, Universit\"at Regensburg, D-93040
Regensburg, Germany}

\begin{abstract}
We address the fundamental interplay between indistinguishability and interactions when discreteness effects are neglected in systems with strictly fixed number of particles. For this end we supplement cluster expansions (many-body canonical techniques where quantum statistics is treated exactly) with short-time/large volume dynamical information where interparticle forces are described non-perturbatively. This approach, specially suitable for the few-body case where it overcomes the inappropriate use of virial expansions, can be consistently combined with scaling considerations, minimal ground-state information and strong coupling expansions in such a way that a single interaction event provides most of the thermodynamic and spectral properties of 1D systems with short range interactions. Our analytical results, in excellent agreement with numerical simulations, show a form of universal integrability of interaction effects for arbitrary confinements.    
\end{abstract}


\maketitle



The description of the physical properties of systems with many (in general interacting) particles is one of the most intriguing and at the same time problematic subjects in modern physics.
As in most cases no exact solutions can be found one falls back either on full numerical simulations or on the problem of identifying simple, basic key features that build up the more complex systems as a whole and their emergent phenomena.
Progress in this direction can be achieved by the combination of quasiparticle, mean field and perturbative methods, with the physical picture corresponding to a system of particle-like excitations evolving under an effective external field and a weak residual interaction~\cite{negele2008,fetter2003}. 

There are basically three reasons for the success of this approach in the past.
First, the previously unsurmountable difficulty in producing high excited states and the consequent focus on ground-state properties where the quasi-particle plus mean-field picture is valid.
Second, the natural interest on extreme regimes where a small parameter can be identified, thus justifying perturbation expansions.
Third, the macroscopically large number of particles typically involved, pushing the system into the limit where well developed grand-canonical methods could be used instead of the fundamental, but far less understood, canonical or microcanonical description.        

The recent experimental realization of quantum systems made of few interacting, identical particles~\cite{greiner2002,serwane2011,preiss2015} and the consequent measurement of their spectral, thermodynamical and dynamical properties poses then a theoretical challenge: while for realistic few-body systems the very concept of mean-field is problematic, the fundamental issue is the lack of analytical tools to describe the interplay between indistinguishability and interaction within a strictly number-constraining formalism.

\begin{figure}[ttt]  
	\includegraphics[width=0.55\columnwidth]{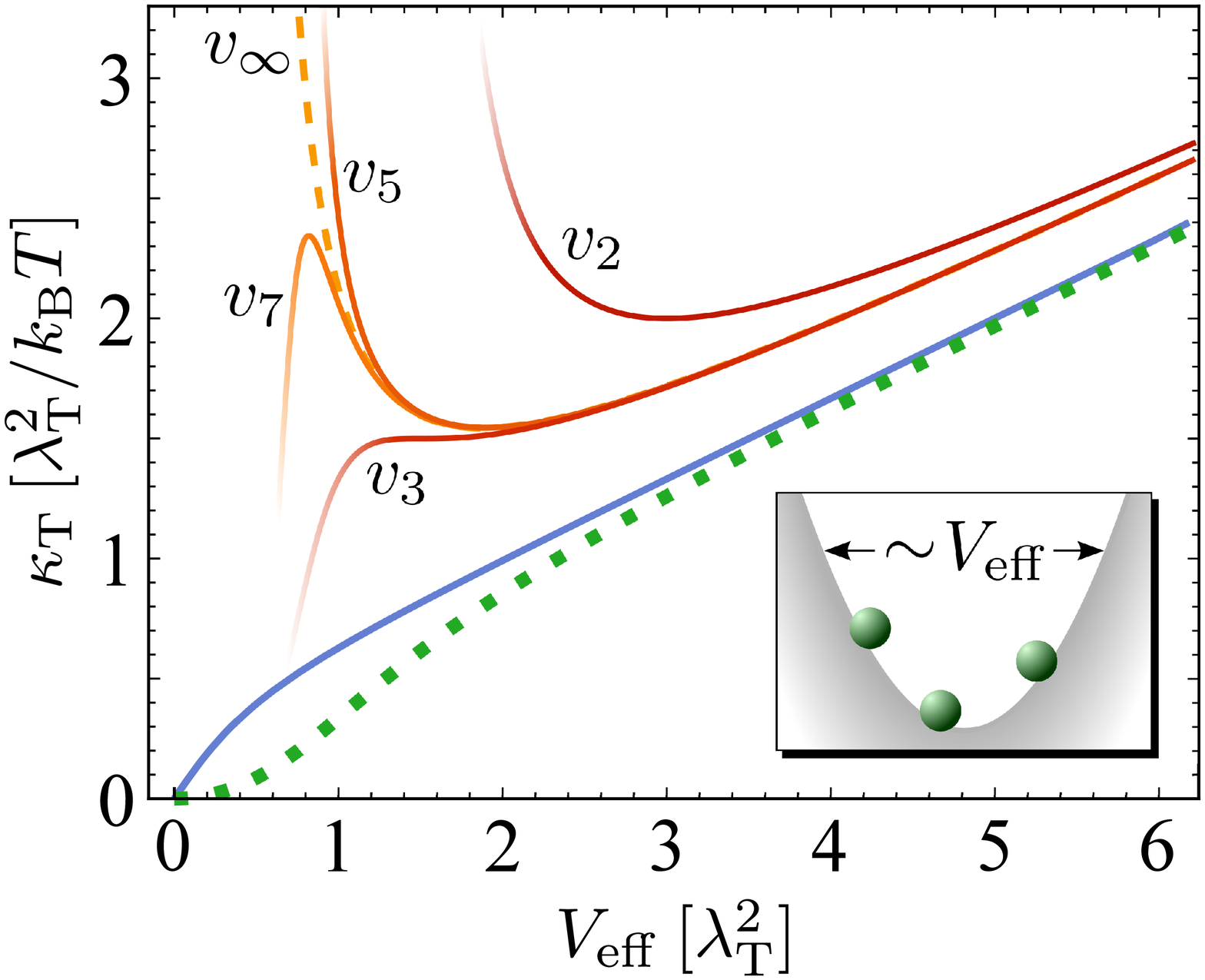}
	\includegraphics[width=0.7\columnwidth]{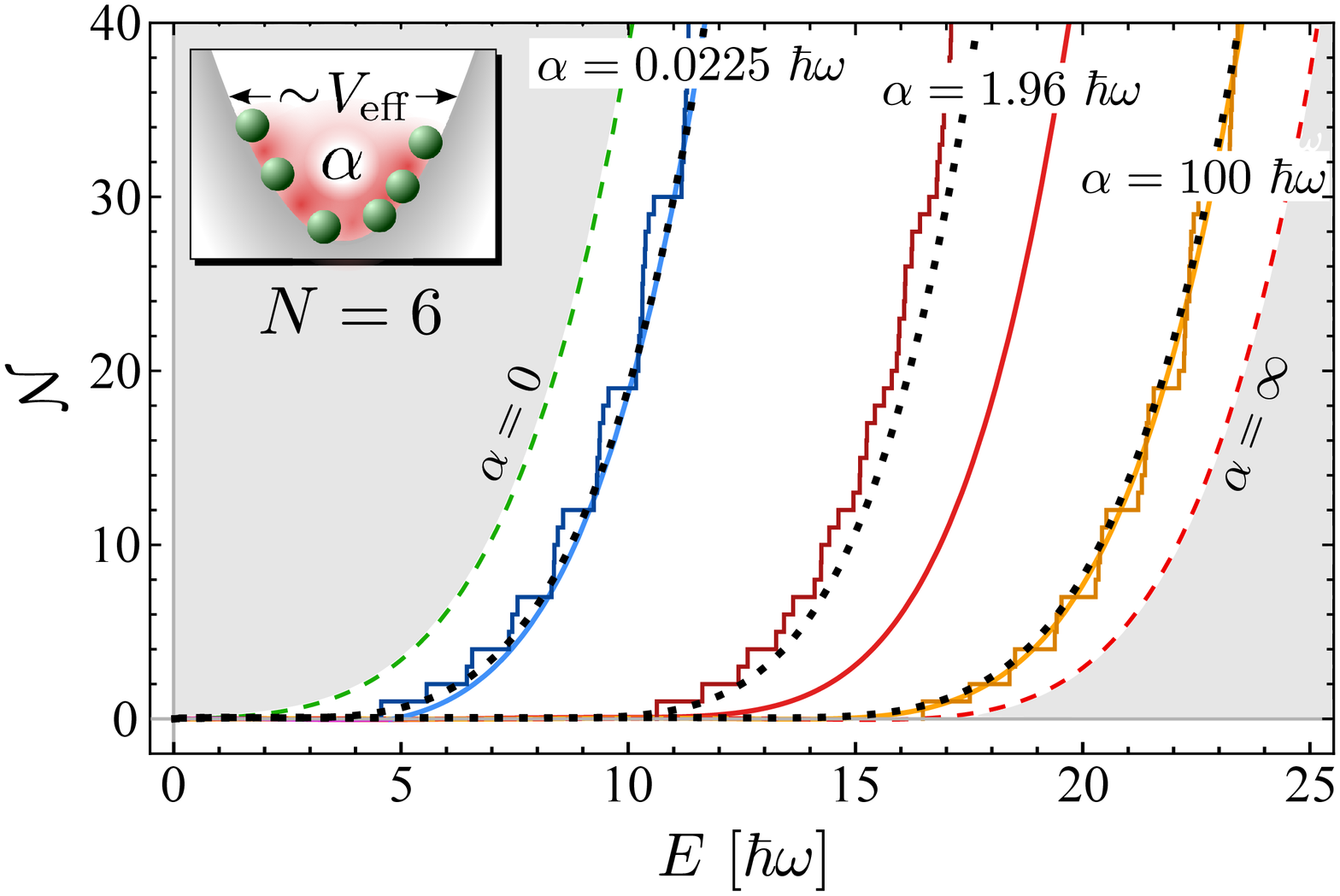}
	\caption{\label{fig:fig1}
		(a) Isothermal compressability of the ideal Bose gas with harmonic confinement and $N=3$ particles.
		While the QCE (solid, blue) using~\eref{eq:Znonint} and~\eref{eq:EOS} fits the numerical values (dotted, green) down to the condensation regime, the virial expansions $v_n$ to various orders $n$ (solid, red-orange tones) give unphysical results.
		Discreteness effects are not taken into account.
		(b) Excitation spectrum of the interacting Bose gas in a harmonic trap ($N=6$) for three values of the scaled interaction strenght $\alpha$.
		Shown is the numerically exact counting function (staircase), and the analytical predictions of our Quantum Cluster Expansion~\eref{eq:rho} (from left to right: weak, strong, strong) directly (solid, smooth), and further combined with scaling considerations that result in the shift method~\eref{eq:Eshift} (dotted). In all cases, two-body interaction events are the whole input of the theory.
	}
\end{figure}

\begin{figure*}[ttt]
		\includegraphics[width=0.9\textwidth]{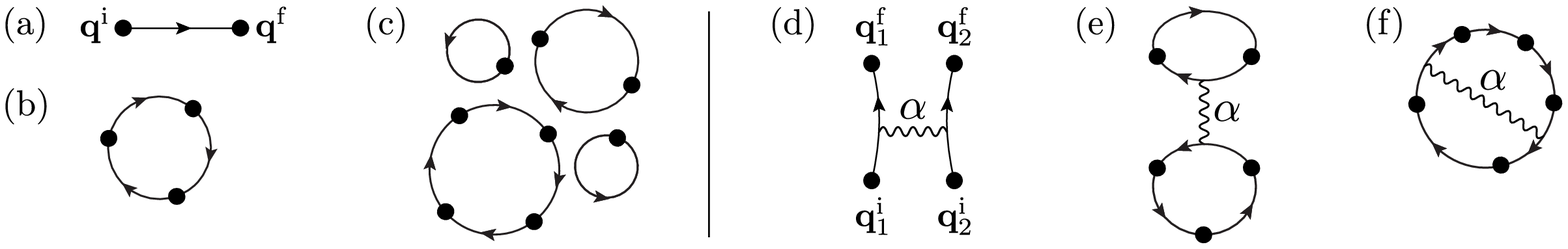}
	\caption{\label{fig:diagrams}(a) SP-propagator $K_0^{(1)}(q^{\rm f}, q^{\rm i};t)$; examples for the contributions (b) $A_{n}$ from a single cycle (here $n=3$) (c) $A_{\mathfrak{N}}$ from a specific clustering (here $\mathfrak{N}=\{1,1,2,4\}, N=8$); (d) interacting part of the two-body propagator $\Delta K^{(2)}((q_1^{\mathrm{f}}, q_2^{\mathrm{i}}),(q_1^{\mathrm{i}},q_2^{\mathrm{i}});t)$; examples for inter-/intra-cycle contributions (e) $A_{(n_1,n_2)}^{\mathrm{inter}}$ and (f) $A_{(n_1,n_2)}^{\mathrm{intra}}$ with $n_1=3, n_2=2$.
	The value of a diagram is defined as the product of all its single-particle and interacting two-body components $K_0^{(1)}$ and $\Delta K^{(2)}$ with all involved (non-terminal) points eventually integrated over the available space. Names of non-terminal points are dropped since they are not an argument of the diagram.}
\end{figure*}

In this paper we fill this gap with a method that provides analytical results (given by sums over a {\it finite} set of diagrams) for spectral and thermodynamical properties of few-body systems where indistinguishability is treated exactly, interactions are treated non-perturbatively, and the total number of particles is strictly fixed.
At the heart of our approach lies the fact that the consistent use of short-time information responsible for the smooth properties of many-body spectra demands that interaction effects are universally given by cluster functions characteristic of quantum integrable models. We further show that this consistence must be applied order by order in the cluster expansion and therefore bounds any extra physical input like scaling considerations, condensation effects and fermionization. In this way most thermodynamic and spectral properties of interacting many-body systems, being quantum integrable or not, turned out to be analytical obtained in terms of a single interaction diagram.

The quality of our approach, the Quantum Cluster Expansion (QCE), is illustrated in \fref{fig:fig1} where we check our results against expensive numerical calculations of thermodynamic and spectral properties of a system made from few interacting bosonic atoms which are harmonically confined, a 1d many-body system of experimental relevance.
As is clearly seen in Fig.~\ref{fig:fig1}a, the failure of grand canonical approaches to describe the thermodynamics in this few-body system is {\it not} just a practical issue: already in the non-interacting limit, including more and more terms of the infinite virial expansions~\cite{gallavotti1999} does not reproduce the correct canonical description. On the contrary, the QCE (here used in its simplest form where condensation effects due to the discreteness of the ground state are not included) provides accurate results down to ultra low temperatures. As shown in Fig.~\ref{fig:fig1}b, when interactions are switched on the consistent combination of one single interaction cluster with scaling and strong coupling expansions provides analytical results for the many-body density of states in excellent agreement with numerical simulations.

The analytical description of few-body systems within the QCE requires two ingredients. First, all information about the discreteness of the many-body spectra is dropped, or included only to account for condensation effects. Like in text-book derivations of grand-canonical potentials for non-interacting systems, this is a standard assumption justified by the high level density of many body systems. Second, interactions will be included only at the pairwise level, but now in a way that is fully consistent with particle exchange symmetry and, importantly, with the short time expansion implicit in the smooth contribution to the spectrum. 

The technical implementation of these assumptions begins with an exact, finite cluster expansion~\cite{ursell1927,kahn1938} of the quantum propagator $K^{(N)}$ for $N$ distinguishable but interacting particles which to first order reads
\begin{eqnarray} \label{eq:QCEK}
		&&K^{(N)}(\bq^\rmf,\bq^\rmi;t) = K^{(N)}_0(\bq^\rmf, \bq^\rmi;t) \nonumber\\
				&&+ \sum_{k<l} K^{(N-2)}_0(\bq^\rmf_{\overline{kl}},\bq^\rmi_{\overline{kl}};t) \Delta K^{(2)}(\bq^\rmf_{kl},\bq^\rmi_{kl};t)
				+ \ldots \,.
\end{eqnarray}
Here $\bq^\rmf$ and $\bq^\rmi$ are the final and initial coordinates, $\overline{kl}$ denotes the set of particle labels excluding $k,l$ and the subscript $0$ refers to non-interacting propagation related through $\Delta K^{(2)}$ with the full two-body propagator by
\begin{equation} \label{eq:K2}
	K^{(2)} = K^{(2)}_0 + \Delta K^{(2)} \,.
\end{equation}
The canonical partition function $Z(\beta)={\rm Tr}K(t = - \rmi \hbar \beta)$ is then given by tracing in the properly (anti)symmetrized coordinate basis, while its inverse Laplace transform
\footnote{We prefer to use the bilateral form of the Laplace transform which gives the correct behavior at negative energies in form of Heaviside-step-functions [\eg $\Linv[\beta^{-1/2}](E)=(\pi E)^{-1/2} \theta(E)$].}
yields the many-body density of states $\varrho(E) = \Linv[Z(\beta)](E)$. Finally, following the approach of~\cite{hur14}, Weyl's method to obtain the smooth single-particle spectrum by replacing the exact quantum propagation for its short-time limit is here generalized to the many-body, interacting case. In the case of $N$ identical non-interacting particles of mass $m$, confined by the homogeneous potential $V(q)=w^{\mu}V(q/w)$ one gets ~\cite{hur14}
\begin{equation} \label{eq:Znonint}
 Z^{(N)}_{0,\pm}(\beta) = \sum_{l=1}^N z_l \left( \frac{V_{\rm eff}}{\lambda_T^d} \right)^l \,, \quad z_l = (\pm 1)^{N-l} C_l^{(N,d)} / l! \,,
\end{equation}
where $\lambda_T = \sqrt{2\pi\hbar^2 \beta / m}$ is the thermal wavelength, the universal constants $C_l^{(N,d)}$ can be found in~\cite{hur14}, and plus~(minus) refer to bosons~(fermions), while the effective dimension $d=D+\frac{2}{\mu}D$ and effective volume
$V_{\rm eff} = (2 \hbar^2 / m e_{0})^{D / \mu} \int \rmd^D\!q \exp({-V({\bf q})/e_{0}})$ with $e_{0}$ an arbitrary energy unit, are given in terms of the physical dimension $D$ and degree of homogeneity $\mu$. The special case of zero external potential is included as $\mu \rightarrow \infty$, $d=D$, and with the available physical volume $V_{\rm eff} = V_D$.

An overview of the possible contributions to the QCE is given by the diagrams shown in Fig.~\ref{fig:diagrams}. The non-interacting part $K^{(N)}_0$ of the propagator factorizes into single-particle (SP) propagators~(see~\fref{fig:diagrams}a) and the contribution to $Z$ corresponding to a permutation $P \in S_N$ is a product of {\it cluster}-contributions, each involving a subset of the particles as large as the cycle-lengths in the cycle-decomposition of $P$.
Let $\mathfrak{N}_P$ denote the multiset with elements $n_i \in \mathbb{N}$ corresponding to the cycle lengths of a permutation $P \in S_N$.
Clearly $\sum_i n_i = N$ and we may use $\mathfrak{N}$ without subscript wherever the assignment is clear from context.
The contribution to the trace of the propagator from one cycle of length $n_i$ is the amplitude~(see~\fref{fig:diagrams}b)
\begin{eqnarray} \label{eq:Ani}
	\mathcal{A}_{n_i}(t) &=& \int \rmd^D\!q_1 \ldots \rmd^D\!q_{n_i} \prod_{k=1}^{n_i} K^{(1)}_0({\bf q}_{k+1},{\bf q}_k; t) \nonumber \\ &=& \int \rmd^D\!q \, K^{(1)}_0({\bf q}, {\bf q};n_i t) \,, 
\end{eqnarray}
where we use the semigroup property of the SP propagator, with the identification ${\bf q}_{n_i+1}:={\bf q}_1$.
Consistently with the short time propagation (as discussed in~\cite{hur14}), we will also use $K_0^{(1)}({\bf q},{\bf q},t) \simeq \rme^{-\frac{i}{\hbar}V({\bf q})t}K_{\rm free}^{(1)}({\bf q},{\bf q},t)$ where $K_{\rm free}$ stands for unconfined propagation.
The full contribution to the non-interacting partition-function corresponding to a permutation is then~(see~\fref{fig:diagrams}c)
\begin{equation}
	\mathcal{A}_\mathfrak{N}(-\rmi \hbar \beta) = \prod_{n \in \mathfrak{N}} \mathcal{A}_{n}(- \rmi \hbar \beta)
\end{equation}
while the partition function is
\begin{equation} \label{eq:Znonintgen}
	Z_{0,\pm}^{(N)}(\beta) = \frac{1}{N!} \sum_{\mathfrak{N} \vdash N} (\pm 1)^{N-l} c^{(N)}_\mathfrak{N} \mathcal{A}_\mathfrak{N}(- \rmi \hbar \beta) \,.
\end{equation}
Here, the sum runs over all partitions $\mathfrak{N}$ of $N$ and
\begin{equation} \label{eq:c}
	c^{(N)}_\mathfrak{N} := \frac{N!}{\prod_{n\in \mathfrak{N}} n \prod_n m_{\mathfrak{N}}(n)!}
\end{equation}
denotes the number of permutations of $N$ with a cycle-decomposition corresponding to $\mathfrak{N}$, where $m_{\mathfrak{N}}(n)$ is the multiplicity of $n$ in $\mathfrak{N}$.
Evaluation of~\eref{eq:Znonintgen} yields then the explicit result~\eref{eq:Znonint}.

At the pairwise level, corresponding to~\eref{eq:QCEK} the effect of interactions is calculated by choosing all possible pairs $\{k,l\}$ of particles and replacing the product $K_0^{(1)}(q_{P(k)},q_k;t) K_0^{(1)}(q_{P(l)},q_l;t)$ in $\mathcal{A}_\mathfrak{N}$ by the interaction term $\Delta K^{(2)}((q_{P(k)},q_{P(l)}),(q_k,q_l);t)$ defined in~\eref{eq:K2}~(see~\fref{fig:diagrams}d).
In the corresponding corrections to $\mathcal{A}_\mathfrak{N}$, the interaction can link two particles involved in either the same or in two different cycles  of $P$, referred as {\it intra-cycle}- and {\it inter-cycle}-contributions respectively.
Basic combinatorics show that the joint contribution to $Z$ from all inter-cycle contributions is 
\begin{equation} \label{eq:Zinter}
	\begin{split}
	Z_{\rm inter}^{(N)} = \left( \begin{matrix} N \\ 2 \end{matrix} \right) \frac{1}{N!} \sum_{n_1=1}^{N-1} \sum_{n_2=1}^{N-n_1} \sum_{\mathfrak{N} \vdash N-n_1-n_2} (\pm 1)^{N-l-2} \\
	\times A_{(n_1,n_2)}^{\rm inter} A_{\mathfrak{N}} \, c_{\mathfrak{N}}^{(N-2)} \,,
	\end{split}
\end{equation}
where $A_{(n_1,n_2)}^{\rm inter}$ is the amplitude of two cycles with $n_1$ and $n_2$ particles involving the interacting part $\Delta K^{(2)}$ for a pair of particles, one of which living in each of the two cycles (see~\fref{fig:diagrams}e). Note that in definition~\eref{eq:c} the cardinality of $\mathfrak{N}$ may differ from the upper index $N$, which is the case in~\eref{eq:Zinter} and that $c_{\mathfrak{N}}^{(N-2)}$ has the meaning of counting the number of permutations of $N$ with two distinct cycles of length $n_1$ and $n_2$ involving particle 1 and 2 respectively and the remaining $N-n_1-n_2$ being composed of cycles with lengths $\mathfrak{N}$. Finally, for the case $n_1+n_2=N$ we consistently define $A_{\{\}}:=1$ and $c_{\{\}}^{(N-2)}:=(N-2)!$. 

For practical use, it is convenient to write~\eref{eq:Zinter} as
\begin{equation}
	Z_{\rm inter}^{(N)} = \frac{1}{2} \sum_{n=2}^{N} (\pm 1)^n Z_{0,\pm}^{(N-n)} \sum_{n_1=1}^{n-1} A_{(n_1,n-n_1)}^{\rm inter} \,,
\end{equation}
which recursively generates $Z_{\rm inter}^{(N)}$ depending on non-interacting partition functions of all smaller particle numbers. Analogue considerations on the intra-cycle contributions $A_{(n_1,n_2)}^{\rm intra}$ shown in~\fref{fig:diagrams}f finally yield the (first order) QCE correction to the partition function
\begin{equation} \label{eq:DeltaZ}
	\Delta Z^{(N)}_\pm = \sum_{n=2}^N (\pm 1)^n Z^{(N-n)}_{0,\pm} \sum_{n_1=1}^{n-1}
		\frac{A_{(n_1,n-n_1)}^{\rm inter} \pm A_{(n_1,n-n_1)}^{\rm intra} }{2}
\end{equation}
and its extension for the case with multiple distinguishable species in~\cite{SM}. 

At the  level of general interactions, equation~(\ref{eq:DeltaZ}) is the main result of this paper. It results from the consistent use of short-time dynamical information, in the spirit of the celebrated Weyl expansion, to obtain thermodynamic and spectral properties that are not sensitive to the discreteness of the many-body spectrum. It is organized in a way such that all contributions coming from indistinguishability are included, while interaction takes place among one pair of particles at the time.

In the following we will show how the QCE can be used to further provide analytical results in situations where the explicit calculation of $A_{(n_1,n_2)}$ is possible. Remarkably, this is the case for the broad case of 1D systems with contact interactions that covers both the integrable Lieb-Liniger model~\cite{lieb1963} as well as the non-integrable and experimentally important case of homogeneous (in particular harmonic) confinement.
  
We choose units by setting $\hbar^2/2m = 1$ such that the thermal wavelength becomes $\lambda_T = \sqrt{4\pi \beta}$.
In these units, the Hamiltonian of the $N$ particle system with coordinates $x_i$ is
\begin{equation} \label{eq:Hdelta}
	\hat{H} =  \sum_{i=1}^N \left(-\frac{\partial^2}{\partial x_i^2} +V(x_{i})\right)+ \sqrt{8 \alpha} \sum_{i<j} \delta(x_i - x_j) \,,
\end{equation}
where $\alpha$ is an energy associated with the strength of the interaction.
\Q{Explain LPA.}
Using the explicit expression for the interacting part of the two-body propagator for this potential~\cite{manoukian1989}, $A_{(n_1,n_2)}^{\rm inter}$ is found to be given by~(\fref{fig:diagrams}e)
\begin{equation} \label{eq:Adelta}
	\begin{split}
	A_{(n_1,n_2)}^{\rm inter} = - \frac{V_{\rm eff}}{\lambda_T^d n^{\frac{d}{2}}} \frac{\sqrt{2\beta \alpha}}{4 \pi }  \int_0^\infty \rmd r \int_{-\infty}^\infty \rmd z \int_0^\infty \rmd u \\
		\times \exp \left[-\frac{1}{8} z^2 - \sqrt{\frac{\beta \alpha}{2}} u - \frac{1}{8}( | \nb z + r | + |r| + u )^2 \right] \,,
	\end{split}
\end{equation}
where $n$ and $\nb$ are related to the numbers of particles involved in the process by
\begin{equation}
	\nb = \sqrt{(2 n_1 n_2 - n_1 - n_2)/n} \,, \quad n=n_1+n_2 \,.
\end{equation}
For $\delta$-interactions it turns out that $A_{(n_1,n_2)}^{\rm intra}=A_{(n_1,n_2)}^{\rm inter}=A_{(n_1,n_2)}$, thus confirming (due to~\eref{eq:DeltaZ}) their vanishing effect on spinless fermions.
Finally, the multiple integrals in~\eref{eq:Adelta} can be reduced by further manipulations to get
\begin{equation} \label{eq:Adelta2}
	\begin{split}
		A_{(n_1,n_2)} =  \frac{V_{\rm eff}}{\lambda_T^d n^{\frac{d}{2}}} \left[ \frac{2}{\pi} \atan \nb - 1 + \frac{2 \nb^2}{\sqrt{\pi(1+\nb^2)}} \sqrt{s} \right.\\
		\left. - \frac{2}{\sqrt{\pi}} \nb \sqrt{s} \rme^s \erfc (\sqrt{s}) + \frac{2}{\sqrt{\pi}} (1 - 2 \nb^2 s) F_{\nb}(s) \right] \,,
	\end{split}
\end{equation}
where we introduced the thermal interaction strength $s=\beta \alpha$.
The remaining integral is defined by
\begin{equation} \label{eq:F}
	F_{\nb}(s) = \rme^{(1+\nb^2)s} \int_0^\infty \rmd z \rme^{-(z-\nb \sqrt{s})^2} \erfc (\sqrt{s} + \nb z) \,,
\end{equation}
and therefore setting $\nb = 0$, $d=1$, and $V_{\rm eff}=L$ recovers the case without confinement involving only two particles $A_{(1,1)} = \frac{L}{\lambda_T} \frac{1}{\sqrt{2}} (-1 + \rme^s \erfc (\sqrt{s}))$~\cite{ghur2015}.
By substituting $F_{\nb}(s)$ into $A_{(n_1,n_2)}$ in Eq.~(\ref{eq:Adelta2}) and using the later into Eq.~(\ref{eq:DeltaZ}) we obtain an analytical expression for the canonical partition function (for specific applications the alternative form of~\eref{eq:F} given in~\cite{SM} improves accuracy in the numerical integration).

Up to this point, we have used the QCE in its simplest form where only one interaction event is taken into count, valid from vanishing to moderately high interaction strength $\alpha$. The quality of the results is, however, drastically extended to higher values of $\alpha$ using the same reduced information by means of a shifting method based on the following universal scaling of $A_{(n_1,n_2)}$ with the (effective) system size, temperature and interaction strength. The main contribution to the $n$-fold integrals in Eq.~(\ref{eq:Ani}) after pairwise replacement with the interacting parts $\Delta K^{(2)}$ comes from the region where all $n$ particles are close to each other, implying fast convergence and allowing us to extend all integrals over relative coordinates to infinity, whereas changes in the center-of-mass are only subject to the external potential, thus yielding the (effective) size $V_{\rm eff}$ of the system as prefactor.
From dimensional analysis the scaling behavior with $V_{\rm eff}/\lambda^{d}_T$ and $\beta \alpha$ then follows as is discussed to more detail in~\cite{SM}. Remarkably, the scaling with $1 / \sqrt{n}$ reminds of the scaling of a non-interacting cycle~\eref{eq:Ani} and fits to the physical picture that all particles that are connected by either symmetry-permutations and/or interaction form a single cluster whose internal wavefunction spreads with a velocity proportional to $n$.

Since the previous arguments hold also for other dimensions (as long as the interaction is ``short-ranged'' in the sense above and can be expressed in terms of a single energy-type parameter $\alpha$ representing its strength) we expect the universal form
\begin{equation} \label{eq:Ascaling}
	A_{(n_1,n_2)}^{\rm inter/intra} = \frac{V_{\rm eff}}{\lambda_T^d} n^{-\frac{d}{2}} a_{(n_1,n_2)}^{\rm inter/intra}(\beta \alpha)
\end{equation}
with the internal part $a_{(n_1,n_2)}^{\rm inter/intra}(s)$ characteristic for the specific interaction and not depending on the external potential.
This scaling together with the scaling of the non-interacting contributions~\eref{eq:Znonint} used in~\eref{eq:DeltaZ} allow us to write the full partition function in first order QCE for $N$ bosons or fermions in a system of (effective) size $V_{\rm eff}$ in the form
\Q{write $\pm$ at $(\Delta)z$?}%
\begin{equation} \label{eq:Z1}
	Z_1^{(N)}(\beta) = \sum_{l=1}^{N} [ z_l + \Delta_1 z_l(\beta \alpha) ] \left( \frac{V_{\rm eff}}{\lambda_T^d} \right)^l \,.
\end{equation}
The interaction-related coefficients are
\Q{write $\pm$ at $(\Delta)z$?}%
\begin{equation} \label{eq:Deltaz}
	\Delta_1 z_l(s) = \sum_{n=2}^{N-l+1} (\pm 1)^n n^{-\frac{d}{2}} z_{l-1}^{(N-n)} \sum_{n_1=1}^{n-1} a_{(n_1,n-n_1)}(s) \,,
\end{equation}
where (defining $z_0^{(m)} := \delta_{m0}$) the general $z$ can be read off from~\eref{eq:Znonint} while $a_{(n_1,n_2)}=(a^{\rm inter}_{(n_1,n_2)} \pm a^{\rm intra}_{(n_1,n_2)})/2$ (given by~\eref{eq:Adelta2} and~\eref{eq:Ascaling} for $\delta$-interactions) must be evaluated for each particular interaction.

We will illustrate the validity of the QCE in general and of the scaling property~\eref{eq:Z1} in particular by comparing its thermodynamical and spectral consequences against numerical simulations.
The QCE mechanical equation of state~\cite{gallavotti1999}
\begin{equation} \label{eq:EOS}
	P(V_{\rm eff},\beta,N,\alpha) = \frac{k_\mathrm{B} T}{V_{\rm eff}} \frac{\sum_{l=1}^N l [z_l + \Delta z_l(\beta \alpha)] \big( \frac{V_{\rm eff}}{\lambda_T^d} \big)^l}
			{\sum_{l=1}^N [z_l + \Delta z_l(\beta \alpha)] \big( \frac{V_{\rm eff}}{\lambda_T^d} \big)^l} \,,
\end{equation}
gives a finite expression for the pressure $P$ in terms of $N$, contrary to virial expansions in the grand canonical treatment, that reproduces very well the exact numerical calculations.
In the same spirit, within QCE the many-body smooth density of states is found to be ($\hbar^2 / 2 m = 1$)
\Q{simplified to solely $f_l$, change SM.}
\begin{equation} \label{eq:rho}
	\bar{\varrho}^{(N)}_\pm(E) = \sum_{l=1}^{N} \left[ \frac{z_l}{\Gamma\!\left(\frac{l d}{2}\right)} +  f_l\!\left(\frac{E}{\alpha}\right) \right]
		\frac{V_{\rm eff}^l E^{\frac{l d}{2}-1} \theta(E)}{(4\pi)^{\frac{l d}{2}}} \, ,
\end{equation}
where the second term between brackets corresponds to the interacting part and the functions $f_l$ can be expressed through elementary functions in the case of $\delta$-interaction~\cite{SM}.
Note that~\eref{eq:rho} shows that the effect of interactions gets suppressed either when the total energy $E \gg \alpha$ or $E \ll \alpha$ for interaction potentials that vanish for  $\alpha \rightarrow 0$ or $\alpha \rightarrow \infty$, respectively.

\begin{figure}[ttt]  
	\includegraphics[width=0.8\columnwidth]{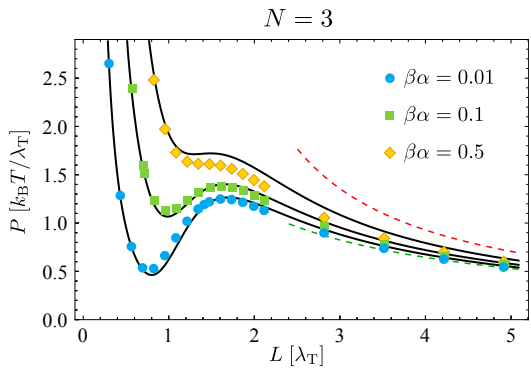}
	\caption{\label{fig:fig4}
		Mechanical equation of state for the three particle Lieb-Liniger model~\cite{lieb1963} with thermal coupling $\beta \alpha = 0.01, 0.1,0.5$.
		Comparison of numerical calculations (symbols) and the analytical QCE~\eref{eq:EOS} combined with minimal analytic information about the lowest two MB states to account for condensation effects~\eref{eq:split}. The system-specific non-monotonous behavior comes from the interplay between repulsive interactions and quasicondensation in the single-particle ground state of zero energy, and is perfectly reproduced by our analytical results.}  
\end{figure}

A special feature of 1D systems with contact interactions that is clearly seen in the numerical results is the apparent mapping linking the limits $\alpha\to 0$ with $\alpha\to \infty$.
This correspondence can be made precise using an exact boson-fermion duality valid for arbitrary $\alpha$~\cite{cheon1999}.
Thus, we can construct the QCE expansion around the strongly interacting regime as an effective spinless fermionic theory, providing again analytical expressions for the partition function~\cite{SM} that perfectly describe the numerical observations in the corresponding regime of large $\alpha$.
In particular, for infinitely strong repulsion, the system behaves as a gas of free fermions in the currently relevant aspects.
For harmonic confinement, fermionization is additionally reflected by the rigid shift $\Delta E_{\infty}$ between the DOS in the two limits.
This feature is incorporated by a suitable generalization/extension of our approach.

Motivated by the general scaling property~\eref{eq:rho} we propose the ansatz
\begin{equation}
\label{eq:Eshift}
\Delta E_{\alpha}=\chi(E/\alpha)\Delta E_{\infty}
\end{equation}
for the energy shift $\Delta E_\alpha$ for finite interaction strength.
In~\cite{SM} we show that the universal scaling $\chi(E/\alpha)$ is uniquely obtained from the first order QCE itself.
It interpolates between the different regimes for $\alpha$ without any fitting.
The shifting method provides again analytical results in good agreement with numerical calculations shown in~\fref{fig:fig1}b.

Besides the possibility of including consistently short-time dynamics into the analytical description given by the QCE when the later is supplemented with scaling considerations and fermionization, a final point is the description of condensation phenomena. Here, and similarly to the usual grand canonical approach, within QCE ultra low temperature effects require that the ground state is treated separately.
Within QCE, this can be achieved by a consistent method where minimal information about the lowest two MB states is combined with the QCE for non-zero temperatures by the ansatz
\begin{equation}
	\label{eq:split}
	Z(\beta) = \rme^{-\beta E_0(V_\mathrm{eff})} + \rme^{-\beta E_1(V_\mathrm{eff})} \sum_{l=0}^N w_l(\beta \alpha) \left( \frac{V_{\rm eff}}{\lambda_T^d} \right)^l \,,
\end{equation}
which can be analytically matched order by order for large $V_\mathrm{eff}$ with~\eref{eq:Z1} to determine the $w_l$ functions.
With this minimal modification, and using only one interaction event, the corresponding modification to~\eref{eq:EOS} shows again excellent agreement with numerical results for the Lieb-Liniger model covering a large regime of interactions and all system sizes~(see~\fref{fig:fig4}).
Moreover, numerical simulations require thousands of many-body energy levels to achieve convergence, and therefore are feasible only because the model at hand is quantum integrable.

Although the QCE exploits the universality of the smooth part of the many-body density of states, in the sense of its dependence with a very restricted set of universal functions together with few geometrical parameters, it can be used to study system specific effects. This is again illustrated in~\fref{fig:fig4} where the non-monotonicity of the pressure as a function of the system's length for three interacting bosons on a ring, a very peculiar consequence of the competition between interactions and bunching, is fully reproduced by our analytical formulas. 

In conclusion, we have shown that the consistent use of short-time/large-volume dynamical information in the description of interacting 1d few-body systems leads to the emergence of robust features depending on a very restricted set of universal functions. In particular, most spectral and thermodynamical observable properties that are not sensitive to the discreteness of the spectrum are resembled by only two-body effects even for non-integrable models.
Our results show that the condition of integrability is too restrictive when one is not interested in the precise form of the many-body spectrum but instead on its smooth part and analytical results can be found for smooth observables for the, previously considered intractable, non-integrable cases.    

We acknowledge financial support from the DPG through the FOR760, and illuminating discussions with Peter Schmelcher, Bruno Eckhart and Benjamin Geiger.
 
\bibliography{qce1d_arxiv}

\cleardoublepage

\newcommand{\sectionQ}[1]{\section{}\vspace*{-10mm}\begin{center}{\bf #1}\end{center}}
\setcounter{section}{1}

\appendix

\sectionQ{Appendix A: Formal derivation of QCE in path-integral formulation}
This section is intended to give analytic support to~\Leref{eq:QCEK}.
To find the first correction to the $N$-body propagator within QCE we start with the exact path-integral representation of the distinguishable propagator
\begin{equation}
	\begin{split}	
		\MoveEqLeft[6] K^{(N)}(\bq^\rmf,\bq^\rmi;t) = \\
		\int_{\bq^\rmi}^{\bq^\rmf} \mathcal{D}\bq(s) 
			&\prod_{k=1}^N \exp \left[{\frac{\rmi}{\hbar}\int_0^t \frac{m}{2} [\dot{\bq}_k(s)]^2 - V_{\rm ext}(\bq_k(s))} \rmd s  \right] \\
		\times &\prod_{k<l} \exp \left[{-\frac{\rmi}{\hbar} \int_0^t V_{\rm int}(\bq_k(s) - \bq_l(s)) \rmd s} \right] \,.
	\end{split}
\end{equation}
Analogous to the Mayer functions in the cluster expansion in classical statistical mechanics we define the \textit{Mayer functionals} $f_{kl}[\bq(s)]$ by
\begin{equation}
	1 + f_{kl}[\bq(s)] := \exp \left[{-\frac{\rmi}{\hbar} \int_0^t V_{\rm int}(\bq_k(s) - \bq_l(s)) \rmd s} \right] \,.
\end{equation}
The next step is to expand the product $\prod_{k<l}$ over pairs into a sum and order its terms by the number of Mayer functionals involved.
\begin{equation}
	\prod_{k<l} (1 + f_{kl}[\bq(s)]) = 1 + \sum_{k<l} f_{kl}[\bq(s)] + \ldots
\end{equation}
In first order QCE we truncate all terms that involve more than one Mayer functional which physically corresponds to neglecting interaction effects that are affecting more than one pair of particles at a time.
Since fo indistinguishable particles all kinds of symmetry related cycle-structures are applied afterwards this will still give non-trivial interaction-induced contributions involving more than two particles.
The first summand involving no Mayer functional gives the non-interacting propagator whereas the next term is evaluated by factorizing the path-integral into independent factors and using the two-body identity
\begin{equation}
	\begin{split}
		&\int_{\bq_{kl}^\rmi}^{\bq_ {kl}^\rmf} \mathcal{D}\bq_{kl}(s)
		\prod_{j = k,l} \exp \left[{\frac{\rmi}{\hbar}\int_0^t \frac{m}{2} [\dot{\bq}_j(s)]^2 - V_{\rm ext}(\bq_j(s))} \rmd s  \right] \\
		&\qquad\qquad\times \left( \exp \left[{-\frac{\rmi}{\hbar} \int_0^t V_{\rm int}(\bq_k(s) - \bq_l(s)) \rmd s} \right] - 1 \right) \\
		&=  K^{(2)}(\bq_{kl}^\rmf, \bq_{kl}^\rmi;t) - K_0^{(1)}(\bq_k^\rmf, \bq_k^\rmi;t) K_0^{(1)}(\bq_l^\rmf, \bq_l^\rmi;t) \\
		&= \Delta K^{(2)}(\bq_{kl}^\rmf, \bq_{kl}^\rmi;t) \,,
	\end{split}
\end{equation}
where $\bq_{kl} = (\bq_k,\bq_l)$, the subscript $0$ denotes propagation amplitudes of the corresponding non-interacting system and hence $\Delta K^{(2)}$ denotes the full interacting part of the two-body propagator $K^{(2)}$.
Together with the $N-2$ independent path-integrals for the remaining particles $j\neq k,l$, which lead to non-interacting single-particle propagators we obtain~\Leref{eq:QCEK}.

\sectionQ{Appendix B: Multiple species}
If multiple distinguishable species of particles are involved the full first order contribution to the overall partition function reads
\begin{equation} \label{sm:eq:DeltaZmultiS}
	\begin{split}
		\Delta Z^{(N_1,\ldots,N_s)} ={} &\sum_{i=1}^s \Delta Z_{\epsilon_i}^{(N_i)} \prod_{j \neq i} Z_{0,\epsilon_j}^{(N_j)} \\
			&{}+ \sum_{i<j} \sum_{n_i=1}^{N_i} \sum_{n_j=1}^{N_j} \epsilon_i^{n_i-1} \epsilon_j^{n_j-1} A_{(n_i,n_j)}^{\rm inter} \\
			&\quad\times Z_{0,\epsilon_i}^{(N_i-n_i)} Z_{0,\epsilon_j}^{(N_j-n_j)} \prod_{k \neq i,j} Z_{0,\epsilon_k}^{(N_k)} \,,
	\end{split}
\end{equation}
where $N_1,\ldots,N_s$ are the numbers of particles in each of the $s$ species and $\epsilon_i = \pm$ reflects the exchange symmetry within species $i$.
The special case $s=2$ of~\eref{sm:eq:DeltaZmultiS} can be used for calculations on the Gaudin-Yang model.
In the given general form,~\eref{sm:eq:DeltaZmultiS} is valid for arbitrary short-ranged interactions addressable with the QCE approach using~\Leref{eq:DeltaZ}.
The interaction-related two-body information then finds its way into~\eref{sm:eq:DeltaZmultiS} through the diagrammatic calculation of $A_{(n_1,n_2)}^{\rm inter}$ and $A_{(n_1,n_2)}^{\rm intra}$ (see~\fref{fig:diagrams}e,f of the Letter) depending on the interacting part of the propagator of two particles living in free space and being subject to the specific interactions (see~\fref{fig:diagrams}d of the Letter).

In the case of $\delta$-interactions the given expression~\Leref{eq:Adelta} can be used in~\eref{sm:eq:DeltaZmultiS}.
Special care has to be taken if some of the particle species are allowed to differ in mass.
Then one has to relax the specific choice of units $\hbar^2 / (2m) =1$ because of ambiguity and take care of the correct masses $m_i$ in all calculations.
This is done by substituting the corresponding thermal de-Broglie wavelength $\lambda_{\rm T} \rightarrow \lambda_{\rm T}^i$ in all expressions involving only one species $i$ on the one hand.
We denote the modified quantities with a tilde and find the two trivial substitutions
\begin{align}
	\tilde{Z}_{0,\epsilon_i} &{}= \left. Z_{0,\epsilon_i} \right|_{\lambda_{\rm T} \rightarrow \lambda_{\rm T}^{i}} \,, \nonumber \\
	\Delta\tilde{Z}_{\epsilon_i} &{}= \left. \Delta Z_{\epsilon_i} \right|_{\lambda_{\rm T} \rightarrow \lambda_{\rm T}^{i}} \,,
\end{align}
with the corresponding thermal de-Broglie wavelength
\begin{equation}
	\lambda_{\rm T}^i = \left( \frac{2 \pi \beta \hbar^2}{m_i} \right)^{\frac{1}{2}} \,.
\end{equation}
On the other hand, the inter-cycle contributions $A_{(n_i,n_j)}^{\rm inter}$ [see~\Leref{eq:Adelta}] between two different species $i$ and $j$ have to be altered by the prescription
\begin{equation}
	\tilde{A}_{(n_i,n_j)}^{\rm inter} = \left( \frac{M_{ij}}{4 \mu_{ij}} \right)^{\frac{1}{2}} \left. A_{(n_i,n_j)}^{\rm inter}
		\right|_{\substack{\lambda_{\rm T} \rightarrow \tilde{\lambda}_{\rm T}^{ij} \\ n \rightarrow \tilde{n}_{ij} \\ \nb \rightarrow \tilde{\nb}_{ij} }} \,,
\end{equation}
where the modified quantities
\begin{align}
	\tilde{\lambda}_{\rm T}^{ij} &{}= \left( \frac{\pi \beta \hbar^2}{\mu_{ij}} \right)^{\frac{1}{2}} \,, \nonumber \\
	\tilde{n}_{ij} &{}= \frac{2 m^{\rm tot}_{ij}}{M_{ij}} \,, \nonumber \\
	\tilde{\nb}_{ij} &{}= \sqrt{\frac{M_{ij}}{m^{\rm tot}_{ij}} n_i n_j - 1} \,,
\end{align}
are defined in terms of the reduced and total mass
\begin{align}
	\mu_{ij} &{}= \frac{m_i m_j}{m_i + m_j} \,, \nonumber \\
	M_{ij} &{}= m_i + m_j
\end{align}
of two representatives of the different species and the total cluster-mass
\begin{equation}
	m^{\rm tot}_{ij} = n_i m_i + n_j m_j \,.
\end{equation}
Naturally, it is also possible to put different interaction-strengths $\alpha_{ij}$ between different species.

\sectionQ{Appendix C: Numerically stable representation of $F_{\nb}(s)$}

The integral given in~\Leref{eq:F} is subject to numerical instability for large values of $s$.
In order to represent the function $F_{\nb}(s)$ in a form where the numerical accuracy is not an essential issue, one can partially treat the integral analytically in a way that the remaining integral gives only small contributions also for large values of $s$.
To acchieve this we first recognize that $F_{\nb}(s)$ can be written in terms of Owen's $T$-function
\begin{equation}
	T(a,b) = \frac{1}{2\pi} \int_0^b \rmd x \frac{\rme^{-\frac{1}{2} a^2 (1+x^2)}}{1+x^2} \,.
\end{equation}
The corresponding expression is
\begin{equation}
	\begin{split}
		F_{\nb}(s) = \rme^{(1+\nb^2)s} &\left[ {\rm erf}(\nb \sqrt{s}) - {\rm erf}(\sqrt{(1+\nb^2)s}) \right. \\
		&\left. {}+ 4 T(\nb \sqrt{2s}, \nb^{-1}) \right] \,,
	\end{split}
\end{equation}
and by use of the general property
\begin{equation}
	T(h,a) + T\!\left(a h, \frac{1}{a}\right) = \frac{1}{4} \left( 1 - {\rm erf}\!\left( \frac{h}{\sqrt{2}} \right) {\rm erf}\!\left( \frac{a h}{\sqrt{2}} \right) \right)
\end{equation}
it is equivalent to
\begin{equation} \label{eqSM:F2}
	\begin{split}
		F_{\nb}(s) = \rme^{(1+\nb^2)s} &\left[ {\rm erfc}(\sqrt{(1+\nb^2)s}) - {\rm erfc}(\nb \sqrt{s}) {\rm erfc}(\sqrt{s}) \right. \\
		&\left. {}+ {\rm erfc}(\sqrt{s}) - 4 T(\sqrt{2 s},\nb) \right] \,.
	\end{split}
\end{equation}
The terms in the first row of this equation are well behaved numerically, since the asymptotics $\rme^{x^2} {\rm erfc}(x) = 1/(\sqrt{\pi} x) + \mathcal{O}(x^{-3})$ for $x\gg1$ are very well known.
The numerical problem now lies in cancellation effects between the two terms of the second row.
To overcome this, we split the Owen $T$ function
\begin{equation}
	4 T(\sqrt{2 s}, \nb) = \frac{2}{\pi} \int_0^\infty \rmd x \frac{\rme^{-s (1+x^2)}}{1+x^2} -
		\frac{2}{\pi} \int_{\nb}^\infty \rmd x \frac{\rme^{-s (1+x^2)}}{1+x^2} \,.
\end{equation}
The first term can be evaluated to
\begin{equation}
	\frac{2}{\pi} \int_0^\infty \rmd x \frac{\rme^{-s (1+x^2)}}{1+x^2} = {\rm erfc}(\sqrt{s}) \,,
\end{equation}
which gets obvious after derivation with respect to $s$, and therefore compensates exactly the term ${\rm erfc}(\sqrt{s})$ in~\eref{eqSM:F2}.
From the remaining integral a factor can be extracted to compensate for the exponential prefactor while keeping it still bounded.
In total one numerically well behaved form of the function $F$ is
\begin{equation}
	\begin{split}
		F_{\nb}(s) ={} &\rme^{(1+\nb^2)s} \left[ {\rm erfc}(\sqrt{(1+\nb^2)s}) - {\rm erfc}(\nb \sqrt{s}) {\rm erfc}(\sqrt{s}) \right] \\
		& {}+ \frac{2}{\pi} \int_{\nb}^\infty \rmd x \frac{\rme^{-s (x^2 - {\nb}^2)}}{1+x^2} \,.
	\end{split}
\end{equation}

\sectionQ{Appendix D: Calculation of QCE contributions}

For comparisons with exact or numerically calculated spectra it is more convenient to use the level counting function $\bar{\mathcal{N}}(E) = \int_{-\infty}^E \rmd E' \bar{\varrho}(E')$ rather than the DOS $\bar{\varrho}(E)$.
Therefore we will give the explicit expressions for the first order QCE-contributions to the coefficients of the former.
One may write
\begin{equation} \label{eq:Calc:N}
	\bar{\mathcal{N}}(E) = \sum_{l=1}^N \left[ \frac{z_l}{\Gamma\left(\frac{l}{2} +1\right)} + g_l^{(N)}\!\left( \frac{E}{\alpha} \right) \right]
		\frac{L^l E^{\frac{l}{2}} \theta(E)}{(4 \pi)^\frac{l}{2}} \,.
\end{equation}
This implies
\begin{equation} \label{sm:eq:ggeneral}
	g_l^{(N)}(\epsilon) = \epsilon^{-\frac{l}{2}} \Linvs \left[ \Delta_1 z_l(s) s^{-\frac{l}{2}-1} \right](\epsilon) \,,
\end{equation}
where the functions $\Delta_1 z_l(s)$ are given by~\Leref{eq:Deltaz}.
The relation to the coefficients of the DOS~[\Leref{eq:rho}] is then given by
\begin{equation}
	\epsilon^{1-\frac{l}{2}} f_l^{(N)}(\epsilon) = \frac{l}{2} g_l^{(N)}(\epsilon) + \epsilon \frac{\rmd}{\rmd \epsilon} g_l^{(N)}(\epsilon) \,.
\end{equation}
For the explicit calculation of~\eref{sm:eq:ggeneral} we split the function
\begin{equation}
	a_{(n_1,n-n_1)}(s) = a_1(s) + a_2(s) + a_3(s) + a_4(s)
\end{equation}
into its four addends
\begin{equation} \label{sm:eq:a1234}
	\begin{split}
		a_1(s) &= \frac{2}{\pi} \atan{\nb} - 1 + \frac{2 \nb^2}{\sqrt{\pi (1+\nb^2)}} \sqrt{s} \,,\\
		a_2(s) &= - \frac{2}{\sqrt{\pi}} \nb \sqrt{s} \rme^s \erfc(\sqrt{s}) \,,\\
		a_3(s) &= \frac{2}{\sqrt{\pi}} F_{\nb}(s) \,,\\
		a_4(s) &= - \frac{4}{\sqrt{\pi}} \nb^2 s F_{\nb}(s) = - 2 \nb^2 s a_3(s) \,,
	\end{split}
\end{equation}
where we have ommitted the dependence on $n_1$ and $n$ through $\nb = \sqrt{{2n_1(n-n_1)}/{n}-1}$ to ease notation.
Together we have
\begin{equation} \label{sm:eq:gfromb}
	g_l^{(N)}(\epsilon) = \sum_{n=2}^{N-l+1} \frac{1}{\sqrt{n}} z_{l-1}^{(N-n)} \sum_{n_1=1}^{n-1} \sum_{j=1}^{4} b_j^{(l)}(\epsilon)
\end{equation}
with
\begin{equation}
	b_j^{(l)}(\epsilon) = \epsilon^{-\frac{l}{2}} \Linvs\left[ s^{-\frac{l}{2}-1} a_j(s) \right](\epsilon) \,.
\end{equation}
In the following explicit expressions for the four $b_j$ are calculated.

\subsection{Calculation of $b_1^{(l)}(\epsilon)$}
Applying standard rules of inverse Laplace transformation to powers of $s$ gives
\begin{equation} \label{sm:eq:Linva1}
	\begin{split}
		b_1^{(l)}(\epsilon) = &\left( \frac{2}{\pi} \atan{\nb} -1 \right) \frac{\theta(\epsilon)}{\Gamma\left( \frac{l}{2}+1 \right)} \\
		& {}+ \frac{2 \nb^2}{\sqrt{\pi (1+\nb^2)}} \frac{\theta(\epsilon)}{\Gamma\left(\frac{l}{2}+\frac{1}{2}\right) \sqrt{\epsilon} } \,.
	\end{split}
\end{equation}
 
\subsection{Calculation of $b_2^{(l)}(\epsilon)$}
Following the recursive approach in~\cite{ghur2015} gives
\begin{equation}
	\begin{split}
		\MoveEqLeft b_2^{(l)}(\epsilon) = - \frac{2 \nb}{\sqrt{\pi}}
			\frac{ \left(1 + \frac{1}{\epsilon}\right)^{\frac{l}{2}-\frac{1}{2}} }{ \Gamma\left(\frac{l}{2}+\frac{1}{2}\right) \sqrt{\epsilon} } h_\lambda(\epsilon) \\
		& {}+ \frac{2 \nb}{\pi} \sum_{k=1}^{\lfloor \frac{l}{2} \rfloor} \frac{\Gamma\left( \frac{l}{2}-k+\frac{1}{2}\right)}
			{ \Gamma\left( \frac{l}{2}-k+1\right) \Gamma\left( \frac{l}{2} + \frac{1}{2} \right) } \left( 1 + \frac{1}{\epsilon}\right)^{k-1} \frac{\theta(\epsilon)}{\epsilon} \,,
	\end{split}
\end{equation}
with the definitions
\begin{equation} \label{sm:eq:h}
	h_\lambda(\epsilon) = \begin{cases}
																\frac{2}{\pi} \theta(\epsilon) \atan ( \sqrt{\epsilon} ) & : \quad \lambda = \frac{1}{2} \,,\\
																\theta(\epsilon) & : \quad \lambda = 0 \,,
	                             \end{cases}
\end{equation}
and
\begin{equation}
	\lambda = \frac{1}{2} ( l\ {\rm mod}\ 2 ) = \begin{cases}
																												\frac{1}{2} & : \quad l\ {\rm odd} \,,\\
																												0 & : \quad l\ {\rm even} \,.\\
	                                                     \end{cases}
\end{equation}
Here $\lfloor q \rfloor$ denotes the integer $n \leq q$ that is closest to $q$.
\\

\subsection{Calculation of $b_3^{(l)}(\epsilon)$}{}
First, we remove the exponential prefactor by defining
\begin{equation}
	\tilde{F}_{\nb}(s) := \rme^{-(1+\nb^2)s} F_{\nb}(s) \,.
\end{equation}
The integral in $\tilde{F}_{\nb}(s)$ can not be evaluated to elementary expressions directly.
In contrast to that its inverse Laplace transform can be related to the solvable derivative given by
\begin{equation} \label{sm:eq:Fprime}
	\rme^{(1+\nb^2)s} \tilde{F}_{\nb}^\prime(s) = \frac{\nb}{2} s^{-\frac{1}{2}} \rme^s \erfc(\sqrt{s}) - \frac{1}{2} \sqrt{1+\nb^2} s^{-\frac{1}{2}} \,.
\end{equation}
Using this observation we calculate
\begin{align}
	\Linvs\left[F_{\nb}(s)\right](\epsilon) &= \Linvs\left[\tilde{F}_{\nb}(s)\right](\epsilon + (1+\nb^2)) \nonumber \\
	&{}= -\frac{ \Linvs\left[\tilde{F}_{\nb}^\prime(s)\right](\epsilon+(1+\nb^2)) }{ \epsilon + (1+\nb^2) } \nonumber \\
	&{}= -\frac{ \Linvs\left[\rme^{(1+\nb^2)s}\tilde{F}_{\nb}^\prime(s)\right](\epsilon) }{ \epsilon + (1+\nb^2) } \nonumber \\
	&{}= (\epsilon + (1+\nb^2))^{-1} \nonumber \\
	&\quad\times\left( \frac{\sqrt{1+\nb^2}}{2\sqrt{\pi}}  \frac{\theta(\epsilon)}{\sqrt{\epsilon}}
		- \frac{\nb}{2\sqrt{\pi}} \frac{\theta(\epsilon)}{\sqrt{1+\epsilon}} \right) \,.
\end{align}
From there we get
\begin{align} \label{sm:eq:InitInt}
	\Linvs\left[s^{-1}F_{\nb}(s)\right](\epsilon) &= \int_{-\infty}^{\epsilon} \rmd x \Linvs\left[F_{\nb}(s)\right](x) \nonumber \\
	&= \frac{\theta(\epsilon)}{\sqrt{\pi}} \left[ \atan\left( \sqrt{\frac{\epsilon}{1+\nb^2}} \right) \right. \nonumber \\
	& \qquad\quad\left. + \atan\left( \sqrt{\frac{\nb^2}{1+\epsilon}} \right) - \atan \nb \right] \,,
\end{align}
and
\begin{equation} \label{sm:eq:InitHalfInt}
	\begin{split}
		\MoveEqLeft \Linvs\left[s^{-\frac{1}{2}}F_{\nb}(s)\right](\epsilon) \\
		&=\int_{-\infty}^\infty \rmd x \Linvs\left[s^{-\frac{1}{2}}\right](\epsilon - x) \Linvs\left[F_{\nb}(s)\right](x) \\
		&\begin{split}
			{}=\frac{\theta(\epsilon)}{2 \pi} \int_0^{\epsilon} \rmd x \frac{1}{\sqrt{\epsilon-x}} &\left[ \frac{\sqrt{1+\nb^2}}{\sqrt{x}(x+(1+\nb^2))} \right. \\
			&\left. {} - \frac{\nb}{\sqrt{1+x}(x+(1+\nb^2))} \right]
		\end{split} \\
		&{}= \frac{\theta(\epsilon)}{\pi} (\epsilon + (1+\nb^2))^{-\frac{1}{2}} \atan\left( \frac{1}{\nb} \sqrt{1+\frac{1+\nb^2}{\epsilon}} \right) \,.
	\end{split}
\end{equation}
We calculate $\Linvs\left[s^{-n} \tilde{F}_{\nb}(s)\right]$ for larger negative powers of $s$ using a recursive approach, where~\eref{sm:eq:InitInt} and~\eref{sm:eq:InitHalfInt} will serve as initial values.
We define
\begin{equation} \label{sm:eq:DefGn}
	G_n(s) := \Gamma(n) s^{-n} \tilde{F}_{\nb}(s) \,,
\end{equation}
where $n$ may be either integer or half-integer.
Taking the derivative of~\eref{sm:eq:DefGn} with respect to $s$ leads to
\begin{equation}
	G_{n+1}(s) = - \frac{\partial}{\partial s} G_n(s) + \Gamma(n) s^{-n} \tilde{F}_{\nb}^\prime(s) \,,
\end{equation}
which implies the recursion relation
\begin{equation} \label{sm:eq:RecRel}
	\begin{split}
		\Linvs\left[G_{n+1}(s)\right](\epsilon) ={} &\epsilon \Linvs\left[G_n(s)\right](\epsilon) \\
		& {}+ \Gamma(n) \Linvs\left[s^{-n} \tilde{F}_{\nb}^\prime(s)\right](\epsilon)
	\end{split}
\end{equation}
for the inverse Laplace transformed objects, where the initial values $\Linvs\left[G_1(s)\right]$ or $\Linvs\left[G_{\frac{1}{2}}(s)\right]$ are given explicitely by~\eref{sm:eq:InitInt} and~\eref{sm:eq:InitHalfInt}.
The solution to~\eref{sm:eq:RecRel} is either given by
\begin{equation} \label{sm:eq:GsolInt}
	\begin{split}
		\Linvs\left[G_{n+1}(s)\right](\epsilon) ={} &\epsilon^n \Linvs\left[G_1(s)\right](\epsilon) \\
		& {}+ \sum_{k=1}^n \epsilon^{n-k} \Gamma(k) \Linvs\left[s^{-k} \tilde{F}_{\nb}^\prime(s)\right](\epsilon)
	\end{split}
\end{equation}
for integer indexes or by
\begin{align} \label{sm:eq:GsolHalfInt}
		&\Linvs\left[G_{n+\frac{1}{2}}(s)\right](\epsilon) = \epsilon^n \Linvs\left[G_\frac{1}{2}(s)\right](\epsilon) \nonumber\\
		& {}\quad+ \sum_{k=0}^{n-1} \epsilon^{n-1-k} \Gamma\left(k+\frac{1}{2}\right) \Linvs\left[s^{-k-\frac{1}{2}} \tilde{F}_{\nb}^\prime(s)\right](\epsilon)
\end{align}
for half-integer indexes.
In the given form, both solutions~\eref{sm:eq:GsolInt} and~\eref{sm:eq:GsolHalfInt} are valid for $n \in \mathbb{N}_0$.
After reintroducing the exponential prefactor,~\eref{sm:eq:GsolInt} and~\eref{sm:eq:GsolHalfInt} become
\begin{eqnarray}
	\begin{split}
		&\Gamma(n+1) \Linvs\left[s^{-n-1} F_{\nb}(s)\right](\epsilon) \\
		&{}= (\epsilon + (1+\nb^2))^n \Linvs\left[s^{-1} F_{\nb}(s)\right](\epsilon) \\
		&\begin{split}
			\quad {}+ \sum_{k=1}^n &(\epsilon+(1+\nb^2))^{n-k} \Gamma(k) \\
			&\times\Linvs\left[s^{-k} \rme^{(1+\nb^2)s} \tilde{F}_{\nb}^\prime(s)\right](\epsilon) \,,
		\end{split}
	\end{split}
\\\nonumber
\end{eqnarray}
and
\begin{eqnarray}
	\begin{split}
		&\Gamma\!\left( n+\frac{1}{2}\right) \Linvs\left[s^{-n-\frac{1}{2}} F_{\nb}(s)\right](\epsilon) \\
		&\qquad\begin{split}
			&{}= \sqrt{\pi} (\epsilon + (1+\nb^2))^n \Linvs\left[s^{-\frac{1}{2}} F_{\nb}(s)\right](\epsilon) \\
			&\quad\begin{split}
				{}+ \sum_{k=1}^n &(\epsilon+(1+\nb^2))^{n-k} \Gamma\!\left(k-\frac{1}{2}\right) \\
				&\times \Linvs\left[s^{-k+\frac{1}{2}} \rme^{(1+\nb^2)s} \tilde{F}_{\nb}^\prime(s)\right](\epsilon) \,,
			\end{split}
		\end{split}
	\end{split} 
\\\nonumber
\end{eqnarray}
where  $n \in \mathbb{N}_0$.
The remaining step is to calculate $\Linvs\left[s^{-n} \rme^{(1+\nb^2)s} \tilde{F}_{\nb}^\prime(s)\right](\epsilon)$ for $n$ being either integer or half-integer.
Using~\eref{sm:eq:Fprime} leads to
\begin{equation}
	\begin{split}
		\MoveEqLeft[4] \Linvs\left[s^{-n} \rme^{(1+\nb^2)s} \tilde{F}_{\nb}^\prime(s)\right](\epsilon) \\
		{}={} &\frac{\nb}{2} \Linvs\left[ s^{-n-1} \sqrt{s} \erfc(\sqrt{s}) \right](\epsilon) \\
		&{}- \frac{1}{2} \sqrt{1+\nb^2} \Linvs\left[ s^{-n-\frac{1}{2}} \right](\epsilon) \\
		{}={} &- \frac{\sqrt{\pi}}{4} \epsilon^n b_2^{(2n)}(\epsilon)
			- \frac{\sqrt{1+\nb^2}}{2 \Gamma(n+\frac{1}{2})} \epsilon^{n-\frac{1}{2}} \theta(\epsilon) \,.
	\end{split}
\end{equation}

For $l \geq -1$ we get
\begin{widetext}
	\begin{equation}\label{sm:eq:b3}
		\begin{split}
			\MoveEqLeft[6] b_3^{(l)}(\epsilon) = \frac{ \left(1+\frac{1+\nb^2}{\epsilon}\right)^{\frac{l}{2}} }{ \Gamma\!\left( \frac{l}{2} + 1 \right) }
				\left[ t_\lambda(\epsilon) - \frac{1}{\sqrt{\pi}} \sum_{k=1}^{\lceil \frac{l}{2} \rceil}
				\Gamma(k-\lambda) \left(1+\frac{1+\nb^2}{\epsilon}\right)^{\lambda-k} \left( \frac{\sqrt{\pi}}{2} b_2^{(2(k-\lambda))}(\epsilon) + \frac{\sqrt{1+\nb^2}}{\Gamma\!\left(k-\lambda+\frac{1}{2}\right)} \frac{\theta(\epsilon)}{\sqrt{\epsilon}} \right) \right] \,,
		\end{split}
	\end{equation}
\end{widetext}
where $\lceil q \rceil$ denotes the integer $n \geq q$ that is closest to $q$ and the function $t_\lambda$ is defined as
\begin{equation}
	t_\lambda(\epsilon) = \begin{cases}
																\frac{2}{\pi} \theta(\epsilon) \atan\left( \frac{1}{\nb} \sqrt{1+\frac{1+\nb^2}{\epsilon}} \right) & : \lambda = \frac{1}{2} \,, \\
																\frac{2}{\pi} \theta(\epsilon) \left[ \atan\left( \sqrt{\frac{\epsilon}{1+\nb^2}} \right) \right. & \\
																	\qquad \quad \left. {}+ \atan\left( \sqrt{\frac{\nb^2}{1+\epsilon}} \right) - \atan \nb \right] & : \lambda = 0 \,.\\
	                             \end{cases}
\end{equation}

\subsection{Calculation of $b_4^{(l)}(\epsilon)$}

Since~\eref{sm:eq:b3} is not only valid for $l \in \mathbb{N}$ but also for the values $l=-1,0$ we can use the simple relation between $a_3$ and $a_4$~\eref{sm:eq:a1234} to get
\begin{equation}
	b_4^{(l)}(\epsilon) = - 2 \nb^2 \frac{1}{\epsilon} b_3^{(l-2)}(\epsilon)
\end{equation}
for all $l \in \mathbb{N}$.

\sectionQ{Appendix E: QCE in fermionization regime}

For arbitrary interaction strengths $\alpha$ a 1D bosonic system with $\delta$-interaction maps exactly to a spinless fermionic system with an effective attractive 0-range interaction potential~\cite{cheon1999} which will here simply be referred to as the anti-$\delta$-interaction.
In order to apply the first order QCE in the effective fermionic theory we need to derive the two-body propagator for the anti-$\delta$-interaction which can be completely achieved on an abstract level relating it back to the propagator in the $\delta$-interacting system.
First, for any two-body propagator $K$ we define the swapping operation denoted by $\bar{K}$ as
\begin{equation}
	\begin{split}
		&\bar{K}((q_1',q_2'),(q_1,q_2)) \\
		&\;=\begin{cases}
			K((q_1',q_2'),(q_1,q_2)), &\text{for } (q_1-q_2)(q_1'-q_2')>0 ,\\
			-K((q_1',q_2'),(q_1,q_2)), &\text{for } (q_1-q_2)(q_1'-q_2')<0 \,,
		\end{cases}
	\end{split}
\end{equation}
which gives a relative sign inversion when the two particles have to cross each other along any classical path from $(q_1,q_2)$ to $(q_1',q_2')$.
Now consider the interacting propagator $K$ of two distinguishable particles subject to the $\delta$-interaction.
It is built from its symmetric part $K_+$ and its antisymmetric part $K_-$ \wrt to particle exchange,
\begin{equation}
	K = K_+ + K_- \,,
\end{equation}
where $K_+$($K_-$) is defined by all symmetric(antisymmetric) eigenfunctions $\psi_\pm(R,r)$ of the two-body system, where $R,r$ denote center-of-mass and relative coordinates, respectively.
The $\delta$-interaction only has an effect on the symmetric wavefunctions $\psi_+(R,r)$, whereas the antisymmetric ones are unaffected $\psi_-(R,r) = \psi_{0,-}(R,r)$, thus we write
\begin{align}
	K_+ &= K_{0,+} + K_\alpha \,,\\
	K_- &= K_{0,-} \,,
\end{align}
where $K_{0,\pm}$ denotes the (anti)symmetric part of the non-interacting propagator and $K_\alpha$ the modification to the symmetric part due to finite interaction.

For the anti-$\delta$-interaction (which will be denoted by a tilde) the opposite is the case and one has unaffected symmetric wavefunction $\tilde{\psi}_+(R,r) = \psi_{0,+}(R,r)$ whereas the antisymmetric wavefunctions $\tilde{\psi}_-(R,r)$ feel the interaction in form of a jump discontinuity at vanishing relative distance $r$ of the particles.
Because of the exact mapping, those antisymmetric wavefunctions are equivalent with the symmetric ones for the $\delta$-interaction with a conditional sign-inversion
\begin{equation}
	\tilde{\psi}_-(R,r) = \mathrm{sign}(r) \psi_+(R,r) \,.
\end{equation}
This sign-inversion is then reflected in the propagator $\tilde{K}$ of two distinguishable particles being subject to the anti-$\delta$-interaction as
\begin{equation}
	\begin{split}
		\tilde{K} &= K_{0,+} + \bar{K}_+ \\
		&= K_{0,+} + \bar{K}_{0,+} + \bar{K}_\alpha \,.
	\end{split}
\end{equation}
For first order QCE calculations one needs then only the modification $\tilde{K}_\alpha$ of the porpagator due to anti-$\delta$-interaction, thus we write
\begin{equation}
	\begin{split}
		\tilde{K} &= K_0 + \tilde{K}_\alpha \\
		&= K_{0,+} + K_{0,-} + \tilde{K}_\alpha \,,
	\end{split}
\end{equation}
and obtain the final result
\begin{equation} \label{eq:Ferm:prop}
	\tilde{K}_\alpha = \bar{K}_{0,+} + \bar{K}_\alpha - K_{0,-} \,.
\end{equation}
A simple test of this result can be done in the limit $\alpha\rightarrow\infty$ where the symmetric propagator for $\delta$-interaction becomes just the swapped version of the free antisymmetric propagator
\begin{equation}
	K_{0,+} + K_\alpha \xrightarrow[\alpha\rightarrow\infty]{} \bar{K}_{0,-} \,,
\end{equation}
so that 
\begin{equation}
	\tilde{K}_\alpha \xrightarrow[\alpha\rightarrow\infty]{} 0 \,,
\end{equation}
which means the fermionic theory is non-interacting in this limit, which confirm the fermionization effect.

Using the relation~\eref{eq:Ferm:prop} in the calculation of the corresponding QCE diagrams involved in the cluster contribution $\tilde{A}_{(n_1,n-n_1)}(s)$ for the fermionic theory one gets then a replacement of the functions $a_{(n_1,n-n_1)} \mapsto \tilde{a}_{(n_1,n-n_1)}$ given by (see~\eref{sm:eq:a1234} for comparison)
\begin{equation} \label{sm:eq:a1234ferm}
	\begin{split}
		\tilde{a}_1(s) &= -\frac{2}{\pi} \frac{\nb}{1+\nb^2} - \frac{2 \nb^2}{\sqrt{\pi (1+\nb^2)}} \sqrt{s} \,,\\
		\tilde{a}_2(s) &= \frac{2}{\sqrt{\pi}} \nb \sqrt{s} \rme^s \erfc(\sqrt{s}) = - a_2(s) \,,\\
		\tilde{a}_3(s) &= \frac{2}{\sqrt{\pi}} F_{\nb}(s) = a_3(s)\,,\\
		\tilde{a}_4(s) &= \frac{4}{\sqrt{\pi}} \nb^2 s F_{\nb}(s) = - a_4(s) \,,
	\end{split}
\end{equation}
and consequently 
\begin{equation} \label{sm:eq:b1234ferm}
	\begin{split}
		\tilde{b}^{(l)}_1(\epsilon) &= -\frac{2}{\pi} \frac{\nb}{1+\nb^2} \frac{\theta(\epsilon)}{\Gamma(\frac{l}{2}+1)} - \frac{2 \nb^2}{\sqrt{\pi (1+\nb^2)}} \frac{\theta(\epsilon)}{\Gamma(\frac{l}{2}+\frac{1}{2}) \sqrt{\epsilon}} \,,\\
		\tilde{b}^{(l)}_2(\epsilon) &= -b^{(l)}_2(\epsilon) \,,\\
		\tilde{b}^{(l)}_3(\epsilon) &= b^{(l)}_3(\epsilon)\,,\\
		\tilde{b}^{(l)}_4(\epsilon) &= -b^{(l)}_4(\epsilon) \,,
	\end{split}
\end{equation}
which can then be used in~\eref{sm:eq:gfromb} and~\eref{eq:Calc:N} together with the non-interacting fermionic coefficients
\begin{equation}
	\tilde{z}_l^{(n)} = (-1)^{n-l} z_l^{(n)}
\end{equation}
to get the corresponding counting functions for the fermionization regime.

\sectionQ{Appendix F: Universal Scaling}\label{sec:Scaling}

\subsection{Generic scaling of spatial potentials}

Suppose the system under observation with (in total) $D$ spatial degrees of freedom is described by a Hamiltonian
\begin{equation} \label{eq:SC:Ham}
	\hat{H} = \hat{T} + V(\hat{\bf q}) \,,
\end{equation}
where $\hat{T}$ is the kinetic energy and $V$ is a spatial potential energy that can be an external potential as well as an interaction potential affecting different particle coordinates.
Suppose further that the potential $V$ scales with parameter $\alpha$ of unit energy that represents its strength.
Other dimensionless parameters $\boldsymbol{\lambda}$ might also be involved.
Moreover, two further constants $\hbar$ and a mass $m$ are allowed to be arguments of the potential.
In case that more than just one mass are entering the Hamiltonian (\eg different particle species or anisotropic mass) the dependence of $V$ on various masses can be substituted by a dependence on one reference-mass (then simply called $m$) and a number of dimensionless parameters $\boldsymbol{\lambda}$  representing the ratios between the actually participating masses and $m$.
In total, the generic assumption is that one can write
\begin{equation} \label{eq:SC:Vgen}
	V({\bf q}) = V(\alpha,\hbar, m, \boldsymbol{\lambda}, {\bf q}) \,,
\end{equation}
with units $[\alpha]=[E]$, $[\lambda_j]=1$, $[q_i]=[x]$, and $[V({\bf q})]=[E]$.
Exceptions of~\eref{eq:SC:Vgen} are potentials that are homogeneous functions of ${\bf q}$ of degree $-2$, namely the (anisotropic) $\sim 1/q^2$ potential, Dirac-Delta potentials $\sim \delta^{(2)}\!\left(\sum_{ij} a_{ij} \left( \begin{smallmatrix} q_i \\ q_j \end{smallmatrix} \right) \right)$ involving two dimensions, linear combinations of the mentioned, and maybe other more exotic constructions.
The reason for this exception is that those potentials intrinsically are given by dimensionless couplings that cannot be transformed into energy-like couplings $\alpha$ by means of the available constants.
At the same time this means that such potentials yield scale-invariant Hamiltonians which need to be regularized to give them physical meaning.
To acchieve that usually the regularized forms are equipped with a physical parameter of the system that is to be modelled.
This parameter must not be dimensionless and is often given as a bound state energy or a scattering length.
Therefore also those exceptional cases are in their final regularized physically meaningful versions again admitting the form~\eref{eq:SC:Vgen}.

With the form~\eref{eq:SC:Vgen} one can write
\begin{equation}
	V(\alpha, \hbar, m, \boldsymbol{\lambda}, {\bf q}) = \alpha \tilde{V}(\alpha, \hbar, m, \boldsymbol{\lambda}, {\bf q}) \,,
\end{equation}
where $\tilde{V}$ is dimensionless.
Therefore its functional dependence on all arguments must be in a way that the latter are combined to dimensionless quantities.
The unique way (up to a dimensionless factor) to do so is given by the scaled coordinates $(\sqrt{2m \alpha}/\hbar) {\bf q}$ which allows one to write
\begin{equation}
	V(\alpha, \hbar, m, \boldsymbol{\lambda}, {\bf q}) = \alpha \bar{V}\!\left(\boldsymbol{\lambda}, \sqrt{\frac{2 m \alpha}{\hbar^2}} {\bf q} \right) \,,
\end{equation}
where $\bar{V}$ is again dimensionless.
The last step introduces a temperature $T$ and related inverse temperature $\beta = 1/ k_{\rm B} T$ which yields the thermal de-Broglie wavelength \begin{equation}
	\lambda_{\rm T} = \left( \frac{m}{2 \pi \hbar^2 \beta} \right)^{-\frac{1}{2}}
\end{equation}
as a length-scale, which defines
\begin{equation} \label{eq:SC:x}
	{\bf x} := \frac{1}{\lambda_{\rm T}} {\bf q}
\end{equation}
as dimensionless, scaled coordinates.
The final general scaling of the potential in terms of ${\bf x}$ is given by
\begin{equation}
	V(\alpha, \hbar, m, \boldsymbol{\lambda}, {\bf q}) = \alpha U\!\left(\boldsymbol{\lambda}, \sqrt{\beta \alpha} {\bf x}\right) \,.
\end{equation}

\subsection{Generic scaling of the propagator}

In this section we consider the evolution of quantum states in the system given by~\eref{eq:SC:Ham} in imaginary time $t = - i \hbar \beta$.
The corresponding (non-unitary) evolution operator for a fixed {\it relaxation ``time''} $\beta$ is $\rme^{-\beta \hat{H}}$.

Let $\left| {\bf q} \right\rangle$ denote the eigenstates of $\hat{\bf q}$ with eigenvalues ${\bf q}$ normalized as
\begin{equation}
	\left\langle {\bf q}' | {\bf q} \right\rangle = \delta^{(D)}({\bf q}'-{\bf q}) \,,
\end{equation}
and let further denote $| \psi \rangle$ an arbitrary state of the system and $\psi({\bf q}) = \langle {\bf q} | \psi \rangle$ its wavefunction.
The action of the evolution operator is given by the action of its exponent which is (everything non-relativistic)
\begin{equation} \label{eq:SC:betaH}
	\langle {\bf q} | \beta \hat{H} | \psi \rangle = \left[ - \beta \sum_i \frac{\hbar^2}{2 m_i} \nabla_{q,i}^2
		+ \beta \alpha U\!\left( \boldsymbol{\lambda}, \sqrt{\beta \alpha} {\bf x} \right) \right] \psi({\bf q}) \,,
\end{equation}
where the scaled version of the potential energy is used (see last subsection).
Here, $\nabla_{q,i}^2$ are Laplacians with respect to some components of ${\bf q}$.
The different masses $m_i$ can be accounted for by defining a reference mass $m$ and additional dimensionless parameters $\lambda_i = m / m_i$ that might also be arguments to the potential $U$.
In addition we rewrite the Laplacians as derivatives $\nabla_{x,i}^2 = \lambda_{\rm T}^2 \nabla_{q,i}^2 $ with respect to the scaled coordinates $x$~\eref{eq:SC:x}.
Eq.~\eref{eq:SC:betaH} then reads
\begin{equation}
	\begin{split}
		&\langle \lambda_{\rm T} {\bf x} | \beta \hat{H} | \psi \rangle \\
		&\qquad= \left[ - \frac{1}{4\pi} \sum_i \lambda_i \nabla_{x,i}^2
			+ \beta \alpha U\!\left( \boldsymbol{\lambda}, \sqrt{\beta \alpha} {\bf x} \right) \right] \psi( \lambda_{\rm T} {\bf x}) \,.
	\end{split}
\end{equation}
This essentially shows that the evolution operator only involves the scaled coordinates ${\bf x}$, some dimensionless parameters $\boldsymbol{\lambda}$ and the product $\beta \alpha$ of inverse temperature (or relaxation ``time'') and the potential coupling.
We write
\begin{equation} \label{eq:SC:evolution}
	\rme^{-\beta \hat{H}} = \rme^{- \hat{h}(x,\boldsymbol{\lambda},\beta \alpha)} \,.
\end{equation}

In order to express the (imaginary) time evolution of any state in the scaled coordinates ${\bf x}$, we define the position eigenstates associated to the scaled coordinates~\eref{eq:SC:x} as
\begin{equation}
	| {\bf x} \rangle = \lambda_{\rm T}^{\frac{D}{2}} | {\bf q} \rangle \,,
\end{equation}
which satisfy the normalization condition
\begin{equation} \label{eq:SC:xnorm}
	\langle {\bf x}' | {\bf x} \rangle = \lambda_{\rm T}^D \langle {\bf q}' | {\bf q} \rangle = \lambda_{\rm T}^D \delta^{(D)}(\lambda_{\rm T} {\bf x}' - \lambda_{\rm T} {\bf x} ) = \delta^{(D)}({\bf x}' - {\bf x}) \,,
\end{equation}
only dependent on scaled coordinates ${\bf x}, {\bf x}^\prime$.
The evolution of any initial state $| \psi(0) \rangle$ to the corresponding final state $| \psi(\beta) \rangle$ in terms of the scaled positions is then given by
\begin{equation}
	\langle {\bf x}^{\rm f} | \psi(\beta) \rangle = \int \rmd^D x^{\rm i}
		\underbrace{\langle {\bf x}^{\rm f} | \rme^{- \hat{h}(x, \boldsymbol{\lambda}, \beta \alpha)} | {\bf x}^{\rm i} \rangle}_{=: k({\bf x}^{\rm f}, {\bf x}^{\rm i}; \boldsymbol{\lambda}, \beta \alpha)}
		\langle {\bf x}^{\rm i} | \psi(0) \rangle \,.
\end{equation}
The scaled evolution kernel $k$ only depends on ${\bf x}^{\rm f}, {\bf x}^{\rm i}, \boldsymbol{\lambda},$ and $\beta \alpha$ (but not on $\alpha$ or $\beta$ alone) and hence the scaling of the evolution kernel (or propagator) in real coordinates is given by
\begin{align}
	K({\bf q}^{\rm f}, {\bf q}^{\rm i}; \beta) &= \langle {\bf q}^{\rm f} | \rme^{-\beta \hat{H}} | {\bf q}^{\rm i} \rangle \nonumber \\
		&= \lambda_{\rm T}^{-D} k(\lambda_{\rm T}^{-1} {\bf q}^{\rm i}, \lambda_{\rm T}^{-1} {\bf q}^{\rm f};  \boldsymbol{\lambda}, \beta \alpha) \,.
\end{align}

The sole dependence of $k$ on the scaled variables $\bxi$, $\bxf$, and $\beta \alpha$ is due to the scaling of the evolution operator~\eref{eq:SC:evolution} on the one hand and the normalization condition~\eref{eq:SC:xnorm} on the other hand.
The latter is here crucial, which can for example be seen when defining $k$ by its differential equation in ${\bf x}^{\rm i}$ and $\beta \alpha$ together with an {\it initial condition}.
One should distinguish two cases.
i) If the potential energy vanishes for $\alpha \rightarrow 0$, the initial condition can be taken at $\beta_0 = 0$ whereas
ii) in case that the potential energy vanishes for $\alpha  \rightarrow \infty$, the reference point will be $\beta_0 \rightarrow \infty$.
Since the scaled kinetic part does not depend on $\beta$, both cases are then summarized by the initial condition 
\begin{equation} \label{eq:SC:kIC}
	\lim_{\beta \rightarrow \beta_0} k({\bf x}^{\rm f}, {\bf x}^{\rm i}; \boldsymbol{\lambda}, \beta \alpha)
		= \prod_i \frac{1}{\sqrt{\lambda_i}} \exp \left[ -\frac{\pi}{\lambda_i} ({\bf x}_i^{\rm f} - {\bf x}_i^{\rm i})^2 \right] \,.
\end{equation}
The finiteness of the initial condition is thereby guaranteed by the proper normalization of $| {\bf x} \rangle$ states.
Note that the finite width of the gaussian in~\eref{eq:SC:kIC} is not contradictory to the point-like initial condition of the propagator in real coordinates
\begin{equation}
	\lim_{\beta \rightarrow 0} \langle {\bf q}^{\rm f} | \rme^{-\beta \hat{H}} | {\bf q}^{\rm i} \rangle = \delta^{(D)}({\bf q}^{\rm f} - {\bf q}^{\rm i}) \,,
\end{equation}
because the scaling ratio between ${\bf x}$ and ${\bf q}$ vanishes for $\beta \rightarrow 0$.

The specific form of the differential equation for $k(\bxf,\bxi; \blam, \beta \alpha)$ is not important for this argument, rather it is the fact that it is an equation involving only $\bxf, \bxi,$ and $\beta \alpha$ which is crucial here.
Nevertheless for completeness the differential equation will be given in an abstract form here and in an explicit form in the next subsection.
To ease notation we drop the dependence on $\blam$, define the {\it thermal coupling} $s := \beta \alpha$ and write $\hat{h}(x, \boldsymbol{\lambda}, \beta \alpha) = \hat{h}(s)$.
The derivative \wrt $s$ is given by
\begin{equation} \label{eq:SC:SEQkabstract}
	\frac{\partial}{\partial s} k(\bxf, \bxi; s) = \langle \bxf | \left( \sum_{j=0}^\infty \frac{ [-\hat{h}(s), - \frac{\partial \hat{h}}{\partial s}]_j}{(j+1)!} \right)
		\rme^{-\hat{h}(s)} | \bxi \rangle \,,
\end{equation}
with the multiple commutator defined as
\begin{equation}
	[ \hat{A}, \hat{B} ]_j = [ \hat{A}, [ \hat{A}, \hat{B} ]_{j-1} ] \quad \mbox{with} \quad [ \hat{A}, \hat{B} ]_0 := \hat{B} \,.
\end{equation}
Since both $\hat{h}(s)$ as well as $\frac{\partial \hat{h}}{\partial s}$ are built from the operators $\hat{\bx}$ and the corresponding conjugate variables $\hat{\bf k}$ with $[ \hat{k}_a , \hat{x}_b ] = - \rmi \delta_{ab}$, the term in brackets in~\eref{eq:SC:SEQkabstract} acts as a differential operator on $\bx$ that depends on $s$.
Meaning one can write
\begin{equation}
	\frac{\partial}{\partial s} k(\bxf, \bxi; s) = \overset{\rightarrow}{\mathcal{D}_{x^{\rm f}}}(s) k(\bxf, \bxi; s) \,,
\end{equation}
with some differential operator $\overset{\rightarrow}{\mathcal{D}_{x^{\rm f}}}(s)$ acting on $\bxf$.
The differential equation can also be given in the more symmetric form
\begin{equation}
	\left( \frac{\partial}{\partial s} - \frac{1}{2} \overset{\rightarrow}{\mathcal{D}_{x^{\rm f}}}(s)
		- \frac{1}{2} \overset{\rightarrow}{\mathcal{D}^\ast_{x^{\rm i}}}(s) \right)  k(\bxf, \bxi; s) = 0 \,.
\end{equation}

\subsection{Alternative derivation using the Schr\"odinger equation}

In this subsection we give an alternative derivation of the scaling behaviour of the evolution kernel employing the Schr\"odinger equation
\begin{align} \label{eq:SC:SEQ}
	&-\frac{\partial}{\partial \beta} K_\alpha(\bqf,\bqi;\beta) \nonumber \\
	&= \left[ - \frac{\hbar^2}{2m} \sum_i \lambda_i \nabla_{q^{\rmf},i}^2
		+ \alpha U\!\left(\blam,\sqrt{\beta \alpha} \frac{\bqf}{\lamT} \right) \right] K_\alpha(\bqf,\bqi;\beta)
\end{align}
for the propagator $K_\alpha$ in real coordinates.
We switch now to scaled coordinates~\eref{eq:SC:x} and define the scaled kernel as
\begin{equation}
	k_\alpha^{\rm sc}(\bxf,\bxi;\beta \alpha) := \lamT^D K_\alpha(\bqf,\bqi;\beta) \,.
\end{equation}
The derivative \wrt  $\beta$ involves then also the scaled coordinates and the prefactor in the following way
\begin{widetext}
	\begin{equation}
		\frac{\partial}{\partial \beta} K_\alpha(\bqf, \bqi; \beta) = \left[ \lamT^{-D} \alpha \frac{\partial}{\partial s}
			+ \lamT^{-D} \frac{\partial \bxf}{\partial \beta} \cdot \boldsymbol{\nabla}_{x^{\rm f}}
			+ \lamT^{-D} \frac{\partial \bxi}{\partial \beta} \cdot \boldsymbol{\nabla}_{x^{\rm i}}
			+ \frac{\partial \lamT^{-D}}{\partial \beta} \right] k_\alpha^{\rm sc}(\bxf, \bxi; s) \,,
	\end{equation}
\end{widetext}
where we have again introduced the {\it thermal coupling} $s = \beta \alpha$.
Recognizing that
\begin{align}
	\frac{\partial {\bf x}^{\rm i(f)}}{\partial \beta} &{}= -\frac{1}{2 \beta} {\bf x}^{\rm i(f)} \,, \\
	\frac{\partial \lamT^{-D}}{\partial \beta} &{}= - \frac{D}{2 \beta} \lamT^{-D} \,, \\
	\frac{\hbar^2}{2m} \nabla_{q^{\rm f},i}^2 &{}= \frac{1}{4 \pi \beta} \nabla_{x^{\rm f},i}^2 \,,
\end{align}
the Schr\"odinger equation~\eref{eq:SC:SEQ} for the scaled kernel becomes
\begin{widetext}
	\begin{equation} \label{eq:SC:SEQsc}
		\left[ -\frac{1}{4 \pi} \sum_i \lambda_i \nabla_{x^{\rm f},i}^2
			- \frac{1}{2} \bxi \cdot \boldsymbol{\nabla}_{x^{\rm i}}
			- \frac{1}{2} \bxf \cdot \boldsymbol{\nabla}_{x^{\rm f}}
			+ s U\!\left(\blam, \sqrt{s} {\bx} \right)
			+ s \frac{\partial}{\partial s}
			- \frac{D}{2}
			\right] k_\alpha^{\rm sc}(\bxf, \bxi; s) = 0 \,.
	\end{equation}
\end{widetext}
Since the differential operator in~\eref{eq:SC:SEQsc} does not depend explicitely on $\alpha$ and the initial condition
\begin{equation}
	\lim_{s \rightarrow s_0} k({\bf x}^{\rm f}, {\bf x}^{\rm i}; s)
		= \prod_i \frac{1}{\sqrt{\lambda_i}} \exp \left[ -\frac{\pi}{\lambda_i} ({\bf x}_i^{\rm f} - {\bf x}_i^{\rm i})^2 \right] \,,
\end{equation}
where
\begin{equation}
	s_0 =
		\begin{cases}
			0  & \mbox{if } \lim_{\alpha \rightarrow 0} V(\bq) = 0 \,, \\
			\infty & \mbox{if } \lim_{\alpha \rightarrow \infty} V(\bq) = 0 \,,
		\end{cases}
\end{equation}
is also independent of $\alpha$, the scaled evolution kernel is completely defined independently of $\alpha$, meaning there is no explicit dependence on $\alpha$.
Therefore one can ommit the subscript and write the scaling property for the propagator in real coordinates and imaginary time as
\begin{equation} \label{eq:SC:PropScaling}
	K_\alpha({\bf q}^{\rm f}, {\bf q}^{\rm i}; \beta)
		= \lambda_{\rm T}^{-D} k^{\rm sc}(\lambda_{\rm T}^{-1} {\bf q}^{\rm i}, \lambda_{\rm T}^{-1} {\bf q}^{\rm f}; \beta \alpha) \,.
\end{equation}
For simplicity we dropped the dependence on dimensionless parameters $\blam$ in $k^{\rm sc}$ which can always exist implicitely.

In the following section, the scaling property~\eref{eq:SC:PropScaling} will be used to derive universal scaling properties for QCE contributions.

\subsection{Universal scaling properties of QCE including external potentials}

Since~\eref{eq:SC:PropScaling} is a general property regardless of the dimension and explicit form of the potential $V$ it holds also for the propagator of systems of $N$ distinguishable particles where $V(\bq) = \sum_{ij} v_{ij}(\bq_i-\bq_j)$ is an interaction potential relating different particles.
Remarkably this holds also if the interaction is applied only on a subset of particles.
Furthermore, also differences of two propagators where the interaction links different subsets of particles in the two cases are still subject to the general scaling with $\beta \alpha$.
This enables us to write Ursell operators of arbitrary order $n$ as
\begin{equation} \label{eq:SC:ursell}
	U^{(n)}_\alpha (\bqf, \bqi; \beta) = \lamT^{-n D} \tilde{u}^{(n)}(\bxf, \bxi; s) \,.
\end{equation}

Moving to the indistinguishable case, an arbittrary cluster contribution involves a product of Ursell operators and the final configuarions are given as a permutation of the initial configuration $\bqf = P(\bqi)$.
This means the integrand for an arbitrary cluster contribution of $n$ particles is given by a function
\begin{equation} \label{eq:SC:cluster}
	K_\alpha^{(n)}(P(\bq),\bq;\beta) = \lamT^{-n D} \tilde{k}^{(n)}(P(\bx), \bx; s) \,,
\end{equation}
where $K$ stands now for a product of ursell operators~\eref{eq:SC:ursell} and $\tilde{k}$ for its scaled version.
The cluster contribution is then the amplitude
\begin{equation} \label{eq:SC:amplitude}
	\int_\Omega \rmd^{D}q_1 \ldots \int_\Omega \rmd^{D}q_n K_\alpha^{(n)}(P(\bq),\bq;\beta) \,.
\end{equation}

Since we talk about a single cluster, there is only one invariant direction in $\bq$-space of a so constructed integrand, which corresponds to the center-of-mass motion.
Otherwise the integral~\eref{eq:SC:amplitude} would be seperable into distinct cluster-contributions per definition.

If additionally a smooth external confinement potential $V_\mathrm{ext}(\bq) = \sum_i v_\mathrm{ext}(\bq_i)$ is applied, we address it by assuming it to be simultaniously constant for all $n$ involved particles.
This is consistent with the short-time philosophy of the QCE since for short times, the relevant spread of the cluster is small compared to the scale of variations in the external potential.
This assumption separates the amplitude~\eref{eq:SC:amplitude} into an \textit{internal part}
\begin{equation}
	\begin{split}
		&Z_\mathrm{int} = \int \rmd^{D}q_2 \ldots \int \rmd^{D}q_n \\
		&\qquad \times K_{\alpha, \mathrm{int}}^{(n)}(P(({\bf 0},\bq_2,\ldots,\bq_n)),({\bf 0},\bq_2,\ldots,\bq_n);\beta) \,,
	\end{split}
\end{equation}
that fixes one of the coordinates, extends the integration over the others to infinity and assumes zero external potential, and an external part
\begin{equation}
	Z_\mathrm{ext} = \int \rmd^{D}q_1 K_{\mathrm{ext},n}^{(1)}(\bq_1,\bq_1;\beta) \,,
\end{equation}
which corresponds to a single particle feeling the $n$-fold external potential.

For the external potential we may use the generic scaling property again, now introducing a parameter $\alpha_\mathrm{ext}$ to write
\begin{equation}
	V_\mathrm{ext}(\bq_1) = \alpha_\mathrm{ext} U_\mathrm{ext}(\sqrt{\beta \alpha_\mathrm{ext}} \bx_1) \,,
\end{equation}
and consequently
\begin{equation}
	Z_\mathrm{ext} = \int \rmd^{D}x_1 \tilde{k}_{\mathrm{ext},n}^{(1)}(\bx_1, \beta \alpha_\mathrm{ext}) \,.
\end{equation}
The short-time effect of the external potential can thereby be considered as a local phase shift
\begin{equation}
	K_{\mathrm{ext},n}^{(1)}(\bq_1,\bq_1; \beta) \simeq \lamT^{-D} \rme^{-\beta n V_\mathrm{ext}(\bq_1)} \,,
\end{equation}
which leads to
\begin{equation}
	Z_\mathrm{ext} = n^{D/2} \xi(\sqrt{n \beta \alpha_\mathrm{ext}}) \,,
\end{equation}
where
\begin{equation}
	\xi(a) := \int \rmd^{D}y \, \rme^{-a^2 U_\mathrm{ext}(a {\bf y})} \,.
\end{equation}
For homogeneously scaling external potentials this gives
\begin{equation}
	Z_\mathrm{ext} = \left( n^{-d/2} \frac{V_\mathrm{eff}}{\lamT^d} \right) n^{D/2} \,,
\end{equation}
with the definitions of effective dimension $d$ and effective volume $V_\mathrm{eff}$ given in the main text.

Using the scaling~\eref{eq:SC:cluster} for the internal part gives
\begin{equation}
	\begin{split}
		&Z_\mathrm{int}(\beta \alpha) = \int \rmd^{D}x_2 \ldots \rmd^{D}x_n \\
		&\qquad \times \tilde{k}^{(n)}_\mathrm{int}(P(({\bf 0},\bx_2,\ldots,\bx_n)),({\bf 0},\bx_2,\ldots,\bx_n);s) \,,
	\end{split}
\end{equation}
which is a function of only $s=\beta \alpha$ so that we are free to write
\begin{equation}
	Z_\mathrm{int} = n^{D/2} a(\beta \alpha) \,,
\end{equation}
which defines the internal amplitude $a$ and where the prefactor is inspired by the free case and explicit QCE(1) calculations in the case of $\delta$-interaction but can be defined like that in any case.

In total, the contribution~\eref{eq:SC:amplitude} from a specific cluster in homogeneous external potentials has the form
\begin{equation}
	A^{(\mathfrak{C})}_\alpha(\beta) = n^{-d/2} \frac{V_\mathrm{eff}}{\lamT^d} a^{(\mathfrak{C})}(\beta \alpha) \,,
\end{equation}
where $\mathfrak{C}$ denotes a specific internal \textit{cluster-structure} that does not depend on any system parameters.

\sectionQ{Appendix G: Energy shifting method}

For the purpose of this section we write the non-interacting counting function as
\begin{equation} \label{eq:ES:N0}
	\mathcal{N}_0(\tilde{E}) = c_N \tilde{E}^{N d/2} + c_{N-1} \tilde{E}^{(N-1)d/2} + \ldots \,,
\end{equation}
with the scaled total energy
\begin{equation}
	\tilde{E} = E \left( \frac{\hbar^2}{2m} V_\mathrm{eff}^{-2/d} \right)^{-1} \,,
\end{equation}
which is a quantity depending on the (effective) system size.
The observation of full shifts $\Delta E_\infty$ between the limits $\alpha \rightarrow 0$ and $\alpha \rightarrow \infty$ is
\begin{equation}
	\Delta \tilde{E}_\infty = \tilde{a} \mathcal{N}^{(2/d-1)/N} \,,
\end{equation}
with some constant $\tilde{a}$, so that the fully shifted counting function
\begin{equation}
	\mathcal{N}_\infty(\tilde{E}) = \mathcal{N}_0(E-\Delta E_\infty)
\end{equation}
can be expanded in the large $\tilde{E}$ limit (which is at the same time a large $V_\mathrm{eff}$ limit) to
\begin{align}
	&\mathcal{N}_\infty(\tilde{E}) \nonumber\\
	&\quad= c_N \tilde{E}^{N \frac{d}{2}} \left( 1+ \frac{\Delta \tilde{E}_\infty}{\tilde{E}} \right)^{N \frac{d}{2}}
		+ c_{N-1} \tilde{E}^{(N-1)\frac{d}{2}} + \ldots \nonumber \\	
	&\quad= c_N \tilde{E}^{N \frac{d}{2}} + \left( c_{N-1} + N \frac{d}{2} \tilde{a} c_N^{(\frac{2}{d}-1)/N} \right) \tilde{E}^{(N-1)\frac{d}{2}} + \ldots \,.
\end{align}
Thus the leading correction from the shift is of the same order in $\tilde{E}$ as the first sub-leading term in~\eref{eq:ES:N0}, so that it can be matched to the sub-leading term of the non-interacting fermionic counting function, which is simply a negative of the free bosonic term.
We identify
\begin{equation}
	c_{N-1} + N \frac{d}{2} \tilde{a} c_N^{(\frac{2}{d}-1)/N} = - c_{N-1} 
\end{equation}
to fix the constant $\tilde{a}$ for the full shift.

In the case of arbitrary interaction strength $\alpha$, the ansatz is
\begin{equation}
	\Delta \tilde{E}_\alpha = \tilde{\chi}(E/\alpha) \mathcal{N}^{(2/d-1)/N} \,,
\end{equation}
where now one has to do a clear distinction between the variables $\tilde{E}$ (which scales with the volume) and $\frac{E}{\alpha}$ (which scales with the interaction strength).
Meaning, we are free to do an expansion for large $\tilde{E}$ while considering $\frac{E}{\alpha}$ as independent variable.
In other words one can do a large $V_\mathrm{eff}$ expansion to get
\begin{align}
	&\mathcal{N}_\alpha(\tilde{E}) \nonumber\\
	&\quad= c_N \tilde{E}^{N \frac{d}{2}} \left( 1+ \frac{\Delta \tilde{E}_\alpha}{\tilde{E}} \right)^{N \frac{d}{2}}
		+ c_{N-1} \tilde{E}^{(N-1)\frac{d}{2}} + \ldots \nonumber \\	
	&\quad= c_N \tilde{E}^{N \frac{d}{2}} \nonumber \\
	&\qquad+ \left( c_{N-1} + N \frac{d}{2} \tilde{\chi}(E/\alpha) c_N^{(\frac{2}{d}-1)/N} \right) \tilde{E}^{(N-1)\frac{d}{2}} + \ldots \,.
\end{align}
As also the first order QCE correction is in general of the order $\tilde{E}^{(N-1)\frac{d}{2}}$, the shifting fraction $\tilde{\chi}$ can be exactly matched to the later.
In the 1D case with $\delta$-interaction and without external potential this can be expressed in terms of~\eref{eq:Calc:N} as
\begin{equation}
	\chi(E/\alpha) = - \frac{\Gamma(\frac{N+1}{2})}{2 z_{N-1}} g_{N-1}^{(N)}\left(\frac{E}{\alpha}\right) \,,
\end{equation}
where $\chi(E/\alpha) = \tilde{\chi}(E/\alpha) \tilde{a}^{-1}$ is now the unscaled energy shift fraction fulfilling
\begin{equation}
	\Delta E_\alpha = \chi(E/\alpha) \Delta E_\infty \,.
\end{equation}
The corresponding expression for homogeneous external potentials is then related by the general scaling property (see appendix F).

\end{document}